\documentclass[twocolumn,showpacs,aps,superscriptaddress]{revtex4-1}

\usepackage{bm}
\usepackage{dcolumn}

\usepackage{times}
\usepackage{float}
\usepackage{amsmath}
\usepackage{amsthm}
\usepackage{amssymb}
\usepackage{amsbsy}
\usepackage{graphicx}
\usepackage{pstricks}
\usepackage{color}
\usepackage{cancel}
\usepackage{subfigure}
\usepackage{enumerate}
\usepackage{mathrsfs}
\usepackage{multirow}
\usepackage{ulem}
\usepackage{wasysym}
\usepackage{sidecap}

\usepackage[colorinlistoftodos,prependcaption]{todonotes}

\begin{document}
\title{Anisotropic exchange Hamiltonian, magnetic phase diagram and domain inversion of Nd$_2$Zr$_2$O$_7$}
\author{J. Xu}
\altaffiliation{jianhui.xu@helmholtz-berlin.de}
\affiliation{\mbox{Helmholtz-Zentrum Berlin f\"{u}r Materialien und Energie GmbH, Hahn-Meitner-Platz 1, D-14109 Berlin, Germany}}
\affiliation{\mbox{Institut f\"{u}r Festk\"{o}rperphysik, Technische Universit\"{a}t Berlin, Hardenbergstra{\ss}e 36, D-10623 Berlin, Germany}}

\author{Owen Benton}
\altaffiliation{john.benton@riken.jp}
\affiliation{\mbox{RIKEN Center for Emergent Matter Science (CEMS), Wako, Saitama, 351-0198, Japan}}

\author{V. K. Anand}
\thanks{Present Address: \'{E}cole Polytechnique F\'{e}d\'{e}rale de Lausanne (EPFL), CH-1015, Lausanne, Switzerland}
\affiliation{\mbox{Helmholtz-Zentrum Berlin f\"{u}r Materialien und Energie GmbH, Hahn-Meitner Platz 1, D-14109 Berlin, Germany}}

\author{A. T. M. N. Islam}
\affiliation{\mbox{Helmholtz-Zentrum Berlin f\"{u}r Materialien und Energie GmbH, Hahn-Meitner Platz 1, D-14109 Berlin, Germany}}

\author{T. Guidi}
\affiliation{\mbox{ISIS facility, Rutherford Appleton Laboratory, Didcot, OX11 0QX, UK}}

\author{G. Ehlers}
\affiliation{\mbox{Oak Ridge National Laboratory, Oak Ridge, P.O. Box 2008, Tennessee 37831, USA}}

\author{\mbox{E. Feng}}
\affiliation{J\"{u}lich Centre for Neutron Science JCNS, Forschungszentrum J\"{u}lich, Outstation at MLZ, Lichtenbergstr. 1, D-85747 Garching, Germany}

\author{Y. Su}
\affiliation{J\"{u}lich Centre for Neutron Science JCNS, Forschungszentrum J\"{u}lich, Outstation at MLZ, Lichtenbergstr. 1, D-85747 Garching, Germany}

\author{A. Sakai}
\affiliation{Center for Electronic Correlations and Magnetism, Institute of Physics, University of Augsburg, D-86135 Augsburg, Germany}

\author{P. Gegenwart}
\affiliation{Center for Electronic Correlations and Magnetism, Institute of Physics, University of Augsburg, D-86135 Augsburg, Germany}

\author{B. Lake}
\altaffiliation{bella.lake@helmholtz-berlin.de}
\affiliation{\mbox{Helmholtz-Zentrum Berlin f\"{u}r Materialien und Energie GmbH, Hahn-Meitner Platz 1, D-14109 Berlin, Germany}}
\affiliation{\mbox{Institut f\"{u}r Festk\"{o}rperphysik, Technische Universit\"{a}t Berlin, Hardenbergstra{\ss}e 36, D-10623 Berlin, Germany}}

\date{\today}

\begin{abstract}
We present thermodynamic and neutron scattering measurements on the quantum spin ice candidate Nd$_2$Zr$_2$O$_7$. The parameterization of the anisotropic exchange Hamiltonian is refined based on high-energy-resolution inelastic neutron scattering data together with thermodynamic data using linear spin wave theory and numerical linked cluster expansion. Magnetic phase diagrams are calculated using classical Monte Carlo simulations with fields along \mbox{[100]}, \mbox{[110]} and \mbox{[111]} crystallographic directions which agree qualitatively with the experiment. Large hysteresis and irreversibility for \mbox{[111]} is reproduced and the microscopic mechanism is revealed by mean field calculations to be the existence of metastable states and domain inversion. Our results shed light on the explanations of the recently observed dynamical kagome ice in Nd$_2$Zr$_2$O$_7$ in \mbox{[111]} fields.
\end{abstract}

\maketitle

\newcommand\lazro{\mbox{La$_2$Zr$_2$O$_7$}}
\newcommand\ndzro{\mbox{Nd$_2$Zr$_2$O$_7$}}
\newcommand\ndsno{\mbox{Nd$_2$Sn$_2$O$_7$}}
\newcommand\ndhfo{\mbox{Nd$_2$Hf$_2$O$_7$}}
\newcommand\ndt{Nd${}^{3+}$}

\newcommand\tn{$T_\text{N}$}
\newcommand\mueff{$\mu_\text{eff}$}
\newcommand\mub{$\mu_\text{B}$}
\newcommand\cp{$C_\text{p}$}

\newcommand\tjx{\tilde{J}_{\tilde{x}}}
\newcommand\tjy{\tilde{J}_{\tilde{y}}}
\newcommand\tjz{\tilde{J}_{\tilde{z}}}

\section{\label{intro}Introduction}
Recently magnetic pyrochlore oxides have attracted a lot of attention because of geometrical frustration on the network of corner-sharing tetrahedra in the crystal structure \cite{Gardner2010rev}. A rich spectrum of magnetic behaviors have been observed in rare-earth pyrochlores. For example, (Ho/Dy)$_2$Ti$_2$O$_7$ show a classical spin ice state and monopole-like excitations originating from the Ising-anisotropy and effective ferromagnetic spin interactions \cite{Harris1997,Fennell2009,Morris2009}. Classical spin ice can be melted by quantum fluctuations coming from transverse terms in the spin Hamiltonian. This leads to quantum spin ice which is a type of long-sought quantum spin liquid with massive many-body entanglement and fractionalized excitations \cite{Hermele2004,Shannon2012,Savary2012,Gingras2014}. The quantum effect in the classical spin ice is generally ignored but could be significant in pyrochlores with Ising-anisotropic light rare earth elements, e.g. Ce, Pr, Sm, Nd \cite{Onoda2010,Onoda2011,Rau2015}.

Among the candidates for quantum spin ice, Nd-containing pyrochlores are quite special due to the peculiar symmetry of the dipolar-octupolar crystal field ground state doublet of the \ndt\ ion \cite{Huang2014,Li2016,Li2017}. It was shown theoretically that the pseudospin-1/2 Hamiltonian has the form of the $XYZ$ model which supports different types of symmetry-enriched quantum spin ice phases, e.g., dipolar and octupolar quantum spin ice states \cite{Lee2012,Li2016,Li2017,Benton2018}.

On the experimental side, the study of \ndzro\ is rather intensive and fruitful compared to the other Nd-containing pyrochlores \cite{Blote1969,Lutique2003,Hatnean2015,Lhotel2015,Xu2015,Xu2016,Petit2016,
Benton2016,Opherden2017,Xu2018,Lhotel2018,Anand2015,Anand2017,Opherden2018,Bertin2015}. The single-ion crystal field ground state was confirmed to be an Ising-anisotropic dipolar-octuplar doublet and the collective ground state was found to be an ``all-in-all-out'' (AIAO) antiferromagnetic order (N\'{e}el temperature $T_\text{N}\approx0.4$\,K) with persistent spin dynamics \cite{Lhotel2015,Xu2015,Xu2016}. Single crystal inelastic neutron scattering (INS) in zero field indicated dynamical spin ice correlations and quantum moment fragmentation, and a parameterized spin Hamiltonian was proposed \cite{Petit2016, Benton2016}. Very recently, quantum spin-1/2 chains and dynamical kagome ice were detected in \ndzro\ with external fields along \mbox{[110]} and \mbox{[111]} directions and the observed neutron scattering spectra permit a complete determination of the exchange parameters \cite{Xu2018,Lhotel2018}. In addition, the susceptibility and magnetization measurements reveal a field-induced transition (critical field $H_\text{c}\sim0.1$\,T) with a large hysteresis in [111] fields \cite{Opherden2017,Lhotel2015}.

Although the measurements and analyses on \ndzro\ are quite comprehensive, a qualitative understanding of the observations based on a spin Hamiltonian is still lacking \cite{Petit2016,Benton2016,Lhotel2018}. The recently determined exchange parameters still cannot describe the INS data in several aspects (e.g. the transition field and spin excitation energies in fields). In this paper, we first refined the spin Hamiltonian by a comprehensive analysis of the high-energy-resolution INS data and thermodynamic properties of \ndzro\ using linear spin wave theory, high temperature series expansion and numerical linked cluster expansion (NLCE). Second, we calculated the phase diagrams for external fields along \mbox{[100]}, \mbox{[110]} and \mbox{[111]} directions using classical Monte Carlo simulations finding qualitative agreement with experiment. Third, we found that the large hysteresis for the \mbox{[111]} field is caused by an irreversibility of the magnetisation or domain inversion due to the existence of a metastable state based on mean field calculations which is critical for explaining the observed dynamical kagome modes in [111] fields \cite{Lhotel2018}.

\begin{figure*}[t]
\centering
\includegraphics[width=\textwidth]{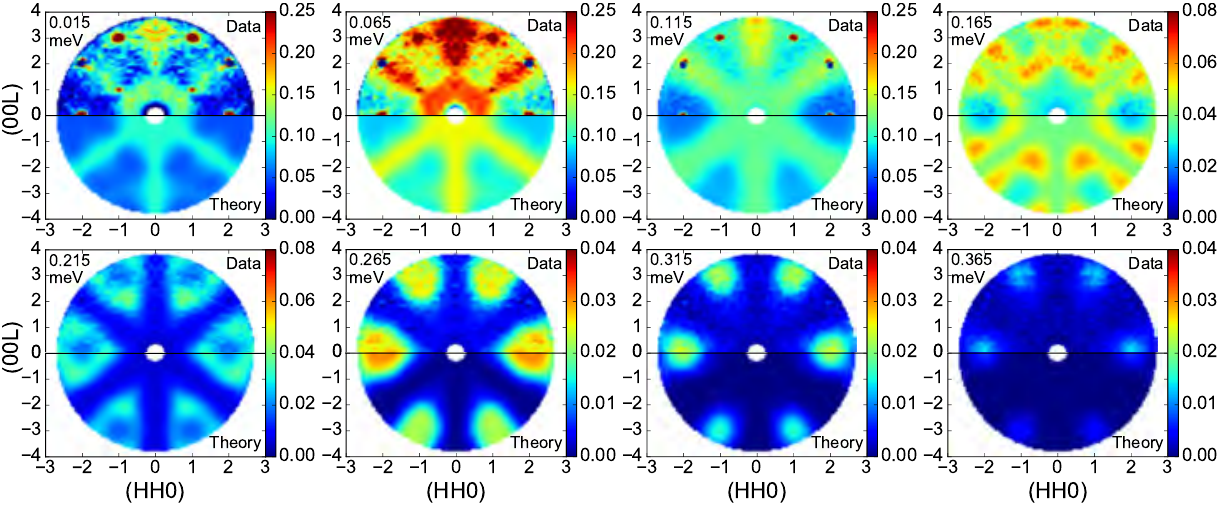}
\caption{Background-subtracted inelastic neutron scattering spectra (in arb. units) measured at 240\,mK on CNCS at SNS at the different constant energies indicated along with corresponding spin wave calculation based on the dipolar-octupolar pseudospin-1/2 model. The data are averaged according to the symmetry of the reciprocal plane. The calculation is convolved with the instrumental resolution of $0.12$\,meV and normalized to the data by an overall factor for a better comparison.
\label{fig:ins_spinw}}
\end{figure*}

\section{\label{methods}Methods}

\subsection{\label{method_sample}{Single crystal growth and structural and magnetic characterizations}}
The \ndzro\ single crystals were grown by using the optical floating zone furnace in the Core Laboratory for Quantum Materials (QM Core Lab) at Helmholtz-Zentrum Berlin (HZB) and characterized by X-ray powder diffraction and Laue diffraction. The susceptibility, magnetization and specific heat measurements above 2\,K were performed on MPMS (SQUID) and PPMS (VSM) (Quantum Design) also at the QM Core Lab at HZB. Specific heat measurements below 1\,K with fields along the \mbox{[111]} crystallographic direction were performed using the thermal relaxation method at Augsburg University  (see Appendixes~\ref{appendix:xrd}, \ref{appendix:chi_mag}, and \ref{appendix:cp} for more details).

\subsection{\label{method_ins_dns}{Neutron scattering}}
The single crystal inelastic neutron scattering experiments were conducted on the direct-geometry time-of-flight (tof) spectrometer Cold Neutron Chopper Spectrometer (CNCS) at Spallation Neutron Source (SNS) in Oak Ridge National Lab and on the indirect-geometry tof spectrometer Osiris at ISIS Neutron Source in Rutherford Appleton Lab \cite{Ehlers2016,Telling2016}. For the CNCS measurement, a single crystal of $\sim2.5$\,g was mounted on a $^3$He insert which cooled the sample down to 240~mK. Neutrons of incident wavelength 4.98~\AA~(3.315\,meV) were used in the high-flux mode of the instrument (energy resolution $\sim0.1$~meV). Data were collected at 240~mK and 20~K with a 360-degree sample rotation at a step of one degree. The large reciprocal space coverage provides an overview of the spin dynamics. To resolve the spin dynamics better, we performed the experiment on Osiris with a higher energy resolution 25\,$\mu$eV using the PG(200) analyser which analyses scattered neutrons of energy 1.84\,meV. The crystal was mounted onto a dilution refrigerator which cooled the sample down to 30\,mK and data were collected at 30\,mK and 20\,K. The data at 20\,K were used as the background for both experiments. The software packages Dave \cite{dave}, Mantid \cite{mantid} and Horace \cite{horace} were used for data processing.

Polarized neutron diffuse scattering with $Z$-polarization analysis was performed on the DNS diffractometer at FRM2, Munich. The single crystal was mounted on a dilution refrigerator. The incident neutron wavelength used was 3.3\,\AA. Data were collected at $\sim300$\,mK and 23\,K (background reference), rotating the sample through 160 degrees in steps of 1 degree.

\subsection{\label{method_simulations}{Spin wave and thermodynamic property simulations}}
The spin waves are calculated based on the linear spin wave theory with Holstein-Primakoff and Bogoliubov transformations \cite{Benton2016}. The Matlab package SpinW is also used for the spin wave calculation \cite{spinw}.

Susceptibility and specific heat were analysed using the numerical linked cluster expansion (NLCE), which is a method of generating a series expansion for quantities in the thermodynamic limit from exact diagonalisation of small clusters. The series is guaranteed to converge at high temperature and for high magnetic fields but can be useful outside these limits (Appendix~\ref{appendix:NLCE}) \cite{Tang2013,Singh2012,Applegate2012}.

The temperature-field phase diagram was simulated using the classical Monte Carlo method. The simulation is done on a cubic cluster of spins with periodic boundary conditions and the spins are treated as classical vectors of fixed length $|\bm{\tau}|=1/2$ (see Appendix~\ref{appendix:MC} for details).

\section{\label{results}Results of inelastic neutron scattering}

\begin{figure}[!htb]
\centering
\includegraphics[width=\linewidth]{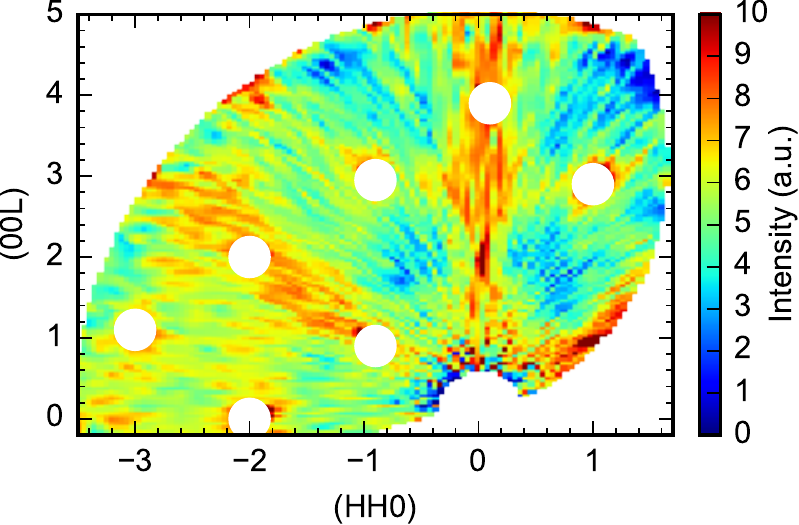}
\caption{Energy-integrated polarized neutron diffuse scattering at $\sim300$\,mK in the $Z$-spin-flip channel, measured in the DNS diffractometer at FRM2.}
\label{fig:dns_sf}
\end{figure}

\begin{figure}[!htb]
\centering
\includegraphics[width=\linewidth]{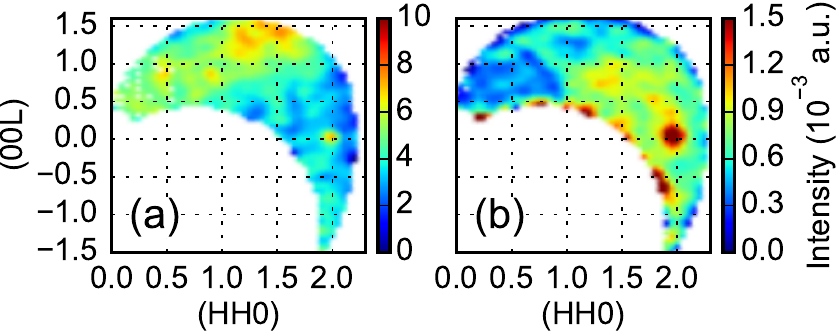}
\caption{The [HHL] plane of the Osiris data integrated over energy ranges (a) [0.06, 0.09] and (b) [0.09, 0.3]~meV which show the pinch point at (111) due to the gapped divergence-free flat modes and orthogonal complementary dispersing pinch point pattern from the curl-free modes of the magnon excitations. The strong intensity at (220) is an instrumental spurion resulted from leakage beyond the elastic channel.}
\label{fig:pinch_moon}
\end{figure}

Figure~\ref{fig:ins_spinw} shows colour-coded intensity maps of constant energy slices through the INS data measured on CNCS in the (HHL) reciprocal plane at 240\,mK in the ordered phase (after background subtraction). Consistent with the data in Ref.~\cite{Petit2016}, we see a highly structured pattern with the symmetry of the scattering plane which evolves with increasing energy transfer: from a gapped pinch point pattern at low energy to a pattern with intensity at the wavevectors [220] and [113] at high energy. The intensity is strongest at around $0.075$\,meV where a pinch point pattern is observed. The energy-integrated polarized neutron diffraction data (Fig.~\ref{fig:dns_sf}) shows a pinch point pattern in agreement with the INS data and the data in Ref.~\cite{Petit2016}. This pinch point pattern is gapped which is clearly shown in Ref.~\cite{Petit2016} and in our high-resolution data measured on Osiris at 30\,mK (Fig.~\ref{fig:pinch_moon} and \ref{fig:fit_3e}). The gapped pinch point pattern is related to divergence-free spin dynamics \cite{Benton2016,Yan2018,Tomonari2018}.

Above the energy of the pinch point mode, there are modes with dispersions starting from the arms of the pinch point pattern and ending at the wavevectors of the AIAO Bragg peaks at $\sim0.27\,$meV. These dispersing modes are recently recognized as dispersing pinch points due to curl-free spin dynamics \cite{Yan2018,Tomonari2018}. An integration of the INS data over energy range [0.09, 0.30]\,meV above the flat pinch point mode, also shows a pinch point pattern with a geometry orthogonal to the flat pinch point pattern at $\sim0.075$\,meV (Fig.~\ref{fig:pinch_moon}). This confirms the theoretical prediction \cite{Yan2018,Tomonari2018}.

\section{\label{sec:spinH}Spin Hamiltonian and spin wave simulations}

The symmetry-allowed nearest-neighbor spin Hamiltonian for the dipolar-octupolar doublet on the pyrochlore lattice is given as \cite{Huang2014}
\begin{equation}
{\cal H}_\text{ex} = \sum_{\langle ij \rangle}J_x\tau^x_i\tau^x_j  +  J_y\tau^y_i\tau^y_j + J_z\tau^z_i\tau^z_j + J_{xz}(\tau^x_i\tau^z_j+\tau^z_i\tau^x_j),
\label{eq:spinH}
\end{equation}
where $\tau^\alpha_i$ is the $\alpha$ component of the pseudospin-1/2 at site $i$ and $J_\alpha$ is the corresponding nearest-neighbor exchange constant ($\alpha=x,y,z$). The pseudospins are defined in the local coordinate frames with the local [111] crystallographic directions as the $z$ axes \cite{Huang2014,Benton2016}. Due to the peculiar symmetry of the dipolar-octupolar doublet, the interactions are uniform for every bond and the cross-coupling $J_{xz}$ can be removed by a pseudospin rotation around the local $y$ axes by a angle $\vartheta$ which leads to the $XYZ$ model \cite{Huang2014,Benton2016}
\begin{equation}
{\cal H}_{XYZ}=
\sum_{\langle ij \rangle}
\big[
\tilde{J}_{\tilde{x}} \tilde{\tau}^{\tilde{x}}_i \tilde{\tau}^{\tilde{x}}_j
+
\tilde{J}_{\tilde{y}} \tilde{\tau}^{\tilde{y}}_i \tilde{\tau}^{\tilde{y}}_j
+
\tilde{J}_{\tilde{z}} \tilde{\tau}^{\tilde{z}}_i \tilde{\tau}^{\tilde{z}}_j
\big].
\label{eq:HXYZ}
\end{equation}
The relations between the exchange parameters and the pseudospin $\bm{\tau}_i$ in the original and in the rotated local frames are \cite{Benton2016}:
\begin{equation}
\begin{aligned}
&J_x = \tilde{J}_{\tilde x} \cos^2\vartheta + \tilde{J}_{\tilde z} \sin^2\vartheta, \\
&J_z = \tilde{J}_{\tilde z} \cos^2\vartheta + \tilde{J}_{\tilde x} \sin^2\vartheta, \\
&\tan2\vartheta = \frac{2 J_{xz}}{J_x- J_z},\\
&J_{xz} = (\tilde{J}_{\tilde x}-\tilde{J}_{\tilde z}) \sin\vartheta \cos\vartheta,
\label{eq:spin_rot_j}
\end{aligned}
\end{equation}

\begin{equation}
\begin{aligned}
\tilde{\tau}^{\tilde{x}}_i &= \tau^x_i\cos\vartheta + \tau^z_i\sin\vartheta , \\
\tilde{\tau}^{\tilde{y}}_i &= \tau^y_i, \\
\tilde{\tau}^{\tilde{z}}_i &= \tau^z_i\cos\vartheta - \tau^x_i\sin\vartheta.
\label{eq:spin_rot_t}
\end{aligned}
\end{equation}

\begin{figure*}[!hbt]
\centering
\includegraphics[width=\textwidth]{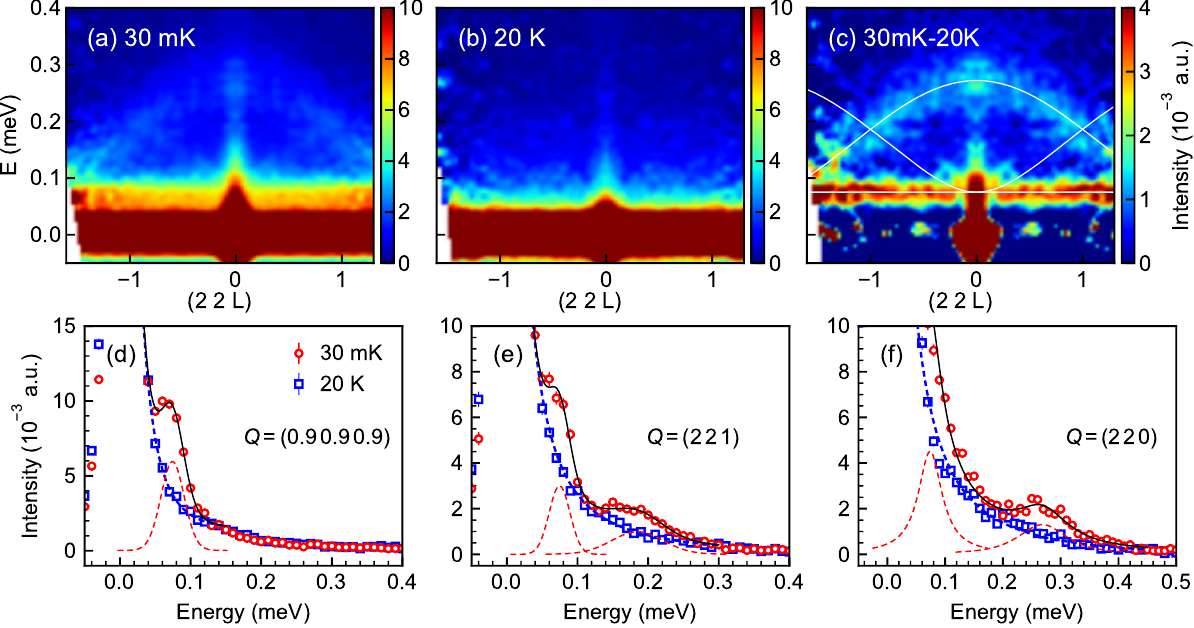}
\caption{High-energy-resolution inelastic neutron scattering data measured on the Osiris spectrometer at ISIS. (a-c): $E$-$Q$ slices along (22L) in reciprocal space at 30\,mK, at 20\,K and with 30mK-20K subtraction. The high intensity at (220) is caused by the strong nuclear and magnetic (below \tn) Bragg peaks. In (c) the spin wave dispersion (white lines) calculated using the refined parameters (Eq.~\ref{eq:j_par}) is plotted over the data. (d-f): $I$-$E$ cuts of the data (markers) at different $Q$ points and fits (black lines) with Gaussian functions (red dashed lines) and estimated background (blue dashed lines, based on the 20\,K data) in order to extract the values of $\Delta_1$ at (0.9,0.9,0.9), $\Delta_2$ at (221) and $\Delta_3$ at (220) of the magnon excitations.}
\label{fig:fit_3e}
\end{figure*}

As pointed out in Ref.~\cite{Benton2016}, the phase diagram and spin dynamics in zero field are determined only by the exchange parameters $\{\tjx,\tjy,\tjz\}$ and the rotation angle $\vartheta$ controls how the pseudospin couples to an external probe because the magnetic moment (along the local \mbox{[111]} directions) is given by
\begin{equation}
\begin{aligned}
m_z & = \mu_\text{B} g_{zz} \tau_z, \\
    & = \mu_\text{B} g_{zz} (\tilde{\tau}^{\tilde{z}} \cos\vartheta + \tilde{\tau}^{\tilde{x}} \sin\vartheta),
\end{aligned}
\end{equation}
where $g_{zz}$ is the only non-zero component of the Ising anisotropic $g$ tensor. Although the system has a strong Ising anisotropy (only $g_{zz}$ is non-zero), the exchange interaction is non-Ising. As a result, transverse spin fluctuations (magnons) with respect to the AIAO ordering direction $\tilde{z}$ may exist, and are visible to neutrons as their projections to the local $z$ axes are non-zero \cite{Benton2016}. As such, despite the presence of strong Ising anisotropy in the system, the neutron scattering data could be analyzed within the framework of spin-wave theory. The neutron scattering data in zero magnetic field in arbitrary units can only determine the three exchange parameters. $\vartheta$ can be determined from the overall neutron scattering intensities in absolute units, the Curie-Weiss temperature, the ordered magnetic moment or the field responses of the system (see Sec.~\ref{sec:g_theta}).

Single crystal inelastic neutron scattering data of \ndzro\ show gapped magnon excitations in the AIAO ordered phase. However, due to the low energy scale of the system, it is difficult to determine the magnon dispersions accurately and thus extract the exchange parameters. According to Ref.~\cite{Benton2016}, the gap to the flat mode is given by
\begin{equation}
\Delta_1 = \sqrt{(3|\tilde{J}_{\tilde{z}}| - \tilde{J}_{\tilde{x}})(3|\tilde{J}_{\tilde{z}}| -\tilde{J}_{\tilde{y}})},
\label{eq:e_gap}
\end{equation}
which provides a constraint for the data analysis. We further extracted another two equations for two characteristic energies which have a simple relation with the exchange parameters, namely the energy $\Delta_2$ at Brillouin zone boundary (e.g. $\bm{Q} = [1 1 0], [1 1 2]$ or $[2 2 1]$) and the highest energy of the dispersion $\Delta_3$ at the Brillouin zone center,
\begin{eqnarray}
\label{eq:e_mid}
\Delta_2 & = & \sqrt{(3|\tilde{J}_{\tilde{z}}|+\tilde{J}_{\tilde{x}})(3|\tilde{J}_{\tilde{z}}| +\tilde{J}_{\tilde{y}})}, \\
\label{eq:e_top}
\Delta_3 & = &~3\sqrt{(|\tilde{J}_{\tilde{z}}|+\tilde{J}_{\tilde{x}}) (|\tilde{J}_{\tilde{z}}| +\tilde{J}_{\tilde{y}})}.
\end{eqnarray}

With Eqs.~\ref{eq:e_gap} to \ref{eq:e_top}, it is convenient to extract $\{\tjx,\tjy,\tjz\}$ unambiguously, provided that the three energies can be determined accurately. Fitting our high-energy-resolution data measured on Osiris (Fig.~\ref{fig:fit_3e}), we have
\begin{equation}
\Delta_1= 0.075(4), \Delta_2=0.186(4), \Delta_3=0.272(5)\,\text{meV}.
\label{eq:deltas}
\end{equation}
Accordingly, the extracted exchange parameters are
\begin{equation}
\tilde{J}_{\tilde{x}}= 0.091(9), \tilde{J}_{\tilde{y}}= 0.014(6), \tilde{J}_{\tilde{z}}= -0.046(2)\,\text{meV}.
\label{eq:j_par}
\end{equation}
Another solution is obtained by swapping the values of $\tjx$\ and $\tjy$\ but it produces the wrong neutron scattering intensity. As shown in Fig.~\ref{fig:ins_spinw}, Fig.~\ref{fig:fit_3e}(c) and in Appendix~\ref{appendix:sw_fit}, the calculated spin wave scattering patterns agree well with the data in zero field.

\begin{figure}[!hbt]
\centering
\includegraphics[width=0.8\linewidth]{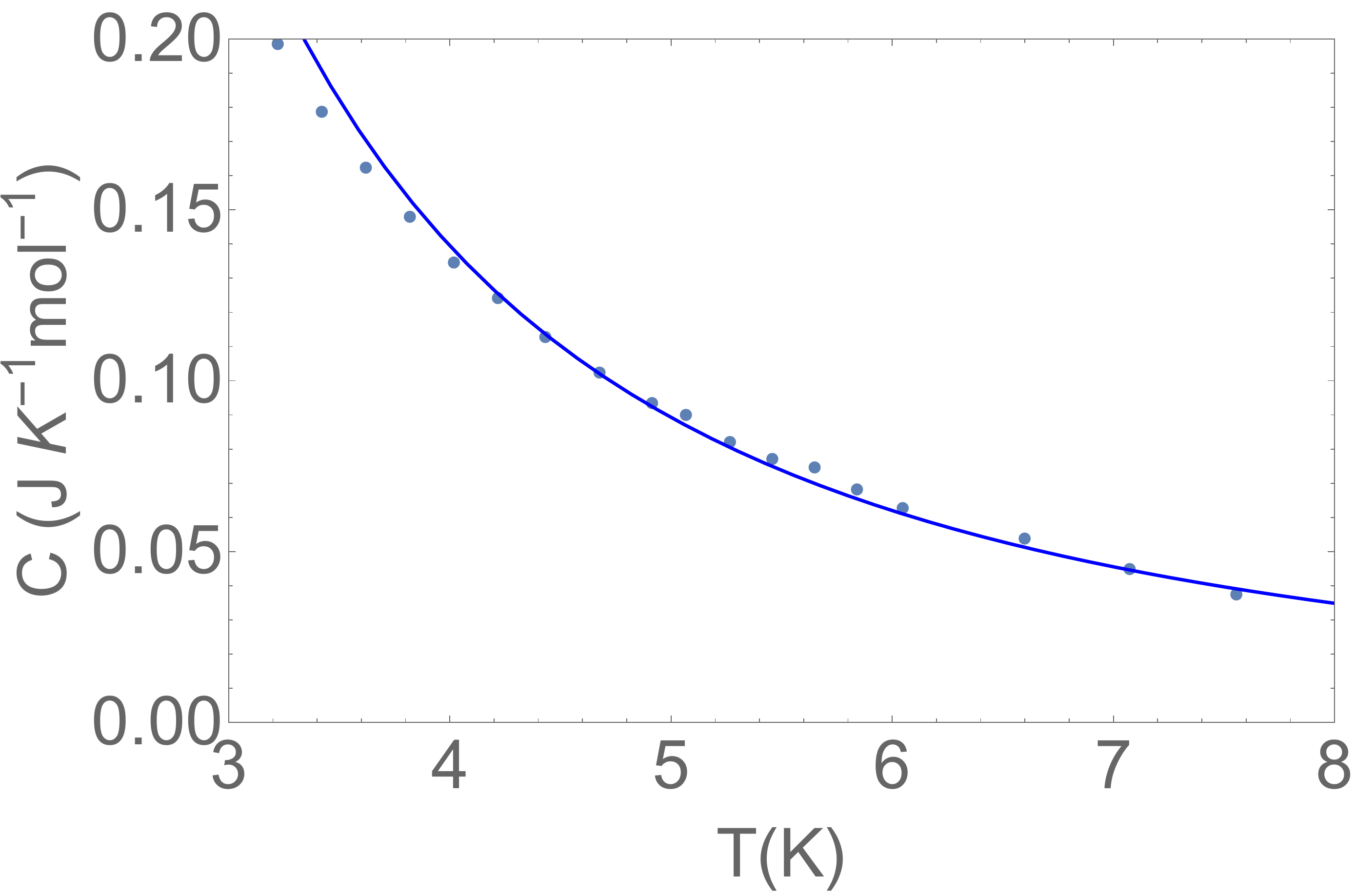}
\caption{Fit of the high temperature $1/T^2$ tail of the heat capacity. The phonon contribution is subtracted which is estimated from fits to the heat capacity of non-magnetic \lazro\ (Appendix~\ref{appendix:cp}). The blue line is a plot of the asymptotic $C_\text{mag}=\alpha/T^2$ law (Eq.~\ref{eq:cp1}) with coefficient $\alpha$ given by Eq.~\ref{eq:cp2} using the refined exchange parameters.}
\label{fig:fit_ht_cp}
\end{figure}

Furthermore, the high-temperature series expansion shows that the high-temperature specific heat can be described by \cite{Baker1967}
\begin{equation}
C_\text{p}\approx\frac{\alpha}{T^2}
\label{eq:cp1}
\end{equation}
with
\begin{equation}
\alpha = \frac{3R}{16k_\text{B}^2} (\tilde{J}_{\tilde{x}}^2+\tilde{J}_{\tilde{y}}^2 + \tilde{J}_{\tilde{z}}^2),
\label{eq:cp2}
\end{equation}
where $R$ is the molar gas constant and $k_\text{B}$ is Boltzmann's constant. As shown in Fig.~\ref{fig:fit_ht_cp}, the calculated heat capacity (without any fitting parameters) agrees well with the experimental data above 4~K (the nuclear lattice contribution has been subtracted) which validates the refined spin Hamiltonian.

\section{\label{sec:g_theta}Determining $g_{zz}$ and $\vartheta$}

The determination of $g_{zz}$ and $\vartheta$ is carried out with $\tilde{J}_{\tilde{x},\tilde{y},\tilde{z}}$ fixed to the values in Eq.~\ref{eq:j_par}. We establish the values of these two parameters by analysing two quantities:
\begin{itemize}
\item the inverse magnetic susceptibility $1/\chi(T)$,
\item the ground state ordered moment $m_\text{ord}$.
\end{itemize}

\begin{figure}[!hbt]
\centering
\includegraphics[width=0.85\linewidth]{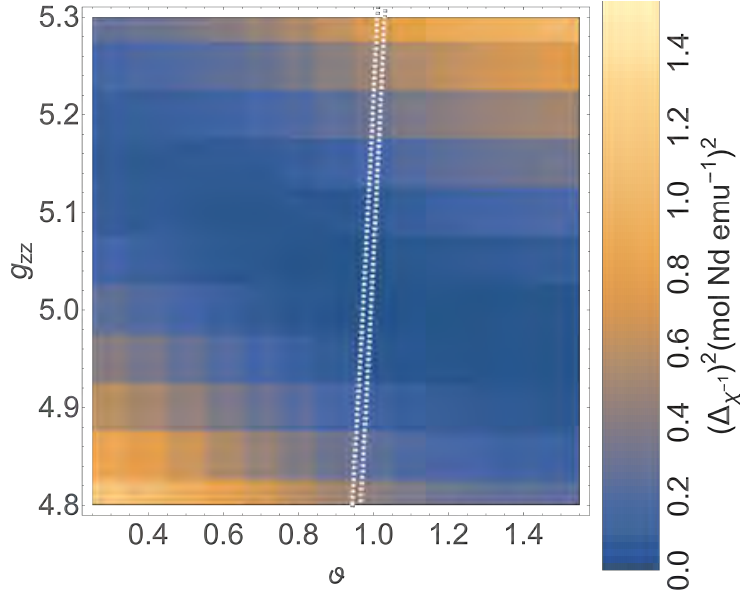}
\caption{Total squared error in the fit to the inverse magnetic susceptibility using third order NLCE as a function of the effective $g$-factor $g_\text{zz}$ and pseudospin rotation angle $\vartheta$. The exchange parameters $\tilde{J}_{x, y, z}$ for the NLCE calculation are set to the refined values (Eq.~\ref{eq:j_par}). The dark blue region indicates an extended region of close agreement between the NLCE calculations and the experimental data. The dashed white lines demark the region where the calculated ordered moment Eq.~\ref{eq:m_ord} agrees with the experimental result $M_\text{ord}=1.26(2)\mu_\text{B}$ from Ref.~\cite{Xu2015}. The overlap between the regions of agreement with the inverse susceptibility and ordered moment gives our estimate of the parameters $g_{zz}$ and $\vartheta$ (Eq.~\ref{eq:g_theta}).}
\label{fig:fit_ht_chi_res}
\end{figure}

The inverse susceptibility is calculated using numerical linked cluster expansion. For the calculation we have used NLCE up to third order. Fig.~\ref{fig:fit_ht_chi_res} shows the total squared error between the experimental and calculated susceptibility as a function of $g_{zz}$ and $\vartheta$ (a constant van Vleck term was included in the fit and optimized independently for each value of $g_{zz}$ and $\vartheta$). There is an extended minimum in the total squared error. To fix the parameters we use a further piece of information: the ground state ordered moment.

The ordered moment for the dipolar AIAO order (with pseudospins oriented  along the local $\tilde{z}$ axes) is given  by
\begin{equation}
m_\text{ord} = \left(\frac{1}{2} - \delta \tilde{\tau}^{\tilde{z}}\right) \cos\vartheta g_{zz} \mu_\text{B},
\label{eq:m_ord}
\end{equation}
where $\delta \tilde{\tau}^{\tilde{z}}$ is the moment reduction from zero point quantum fluctuations. We calculate $\delta \tilde{\tau}^{\tilde{z}}\approx0.05$ from linear spin wave theory using the exchange parameters in Eq.~\ref{eq:j_par}.

\begin{figure}[!hbt]
\centering
\includegraphics[width=\linewidth]{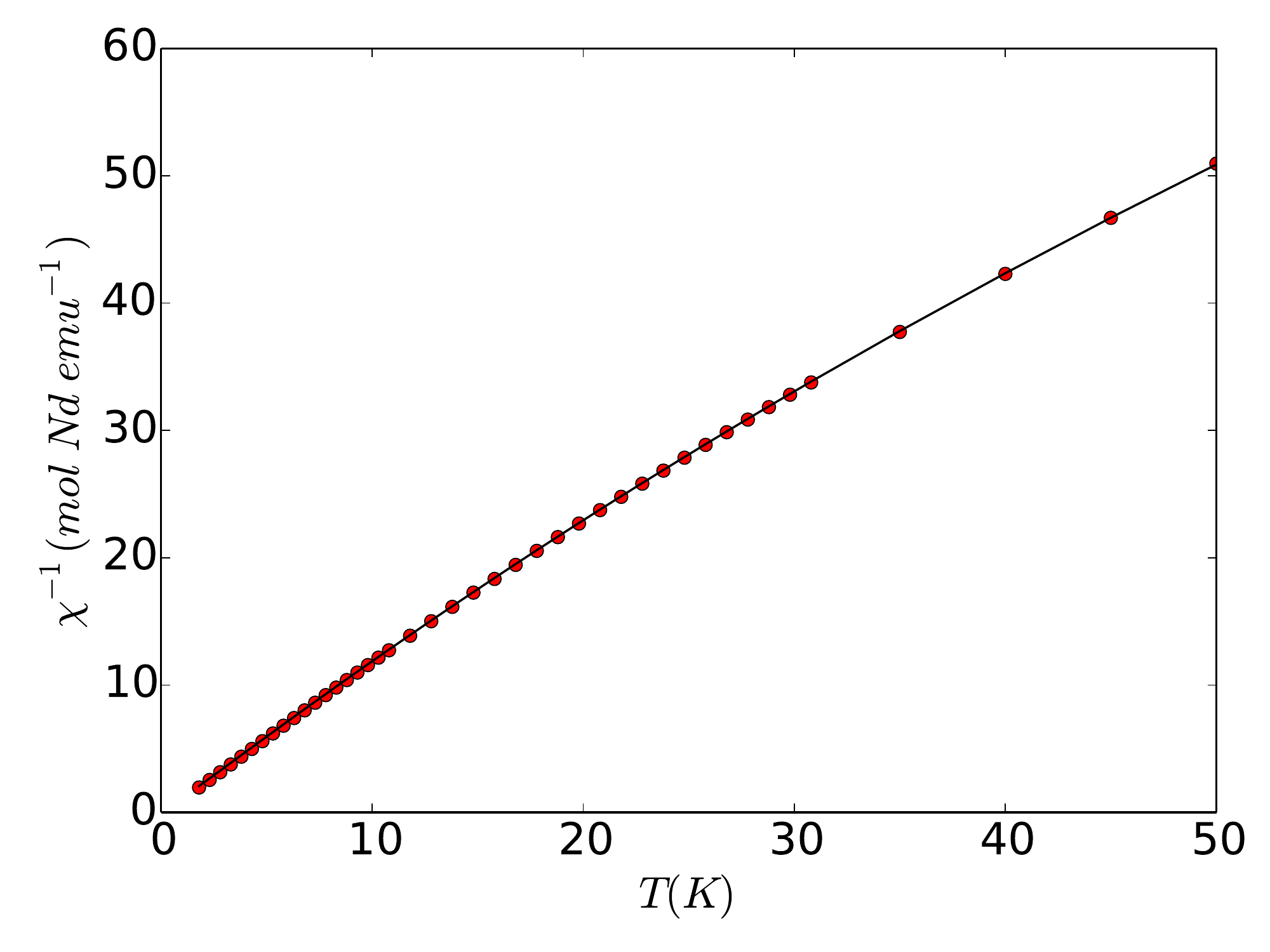}
\caption{Fit to the inverse magnetic susceptibility using third order numerical linked cluster expansion. The parameters $\tjx$, $\tjy$, $\tjz$, $\vartheta$ and $g_{zz}$ are set to the refined values (Eq.~\ref{eq:j_par} and \ref{eq:g_theta}). A constant van Vleck term $3.9\times10^{-3}$\,emu/mol~Nd was included in the fit.}
\label{fig:fit_ht_chi}
\end{figure}

Experimentally, we have the ordered moment at 0.1\,K $m_\text{ord}=1.26(2)\mu_\text{B}$ according the neutron diffraction experiments in Ref.~\cite{Xu2015}. The region of parameter space which is consistent with this is marked by the dashed white lines in Fig.~\ref{fig:fit_ht_chi_res}. The overlap between this region and the extended minimum in the squared error between experimental and theoretical inverse susceptibility gives our estimates:
\begin{equation}
\label{eq:g_theta}
g_{zz}=5.0(1), \vartheta=0.98(3)~ \text{radian}.
\end{equation}
The resulting fit to the inverse susceptibility is shown in Fig.~\ref{fig:fit_ht_chi}. The obtain $g_{zz}$ accounts for the measured saturated magnetization for the single crystal (Appendix~\ref{appendix:chi_mag}). Table~\ref{tab:jpar_compare} presents a comparison of the determined spin Hamiltonian with the ones in previous works. Our parameters are consistent with Ref.~\cite{Lhotel2018} published recently. The major difference comes from $g_{zz}$ and $\vartheta$ which is because a fixed $g_{zz}$ was used in Ref.~\cite{Lhotel2018} rather than being fitted.

\begin{table}
\caption{\label{tab:jpar_compare}Comparison of the Hamiltonian parameterisations and the calculated $T_\text{CW}$ of \ndzro.}
\begin{tabular}{ c c c c c c c}
  \hline
  $\tjx$ (meV) & $\tjy$ (meV) & $\tjz$ (meV)& $\vartheta$ (rad.) & $g_{zz}$ & $T_\text{CW}$(K)   & Refs.\\
  \hline
  -0.047       & 0            & 0.103       & 0                  & 4.5      & -                   & \cite{Petit2016} \\
   0.103       & 0            & -0.047      & 0.83               & -        & -                   & \cite{Benton2016} \\
   0.086(4)    & 0.006(20)    &-0.043(4)    & 1.26(17)           & 4.55     & 0.43(8)             & \cite{Lhotel2018} \\
   0.091(9)    & 0.014(6)     &-0.046(2)    & 0.98(3)            & 5.0(1)   & 0.28(4)             & This work \\
  \hline
\end{tabular}
\end{table}

In addition, the Curie-Weiss (CW) temperature is given in Ref.~\cite{Benton2016}
\begin{equation}
T_\text{CW} = \frac{\tilde{J}_{\tilde z} \cos^2\vartheta + \tilde{J}_{\tilde x} \sin^2\vartheta}{2k_B}.
\label{eq:ins_tcw}
\end{equation}
It yields $T_\text{CW}=0.28(4)$\,K which is consistent with the experimental data 0.31(1)\,K (Appendix~\ref{appendix:chi_mag}), solving the puzzle that \ndzro\ has an antiferromagnetic order but positive Curie-Weiss temperature \cite{Hatnean2015, Lhotel2015, Xu2015, Benton2016}.

Finally, the exchange parameters in the original local coordinate frames are
\begin{equation}
\begin{aligned}
J_x &= -0.0032(48),   &J_y &=     0.014(6), \\
J_z &= 0.049(7),   &J_{xz} &=  0.063(5)\,\text{meV}.
\end{aligned}
\end{equation}

\section{\label{sec:cp_in_fields}Heat capacity in \mbox{[111]} fields}

\begin{figure}[!hbt]
\centering
\includegraphics[width=0.9\linewidth]{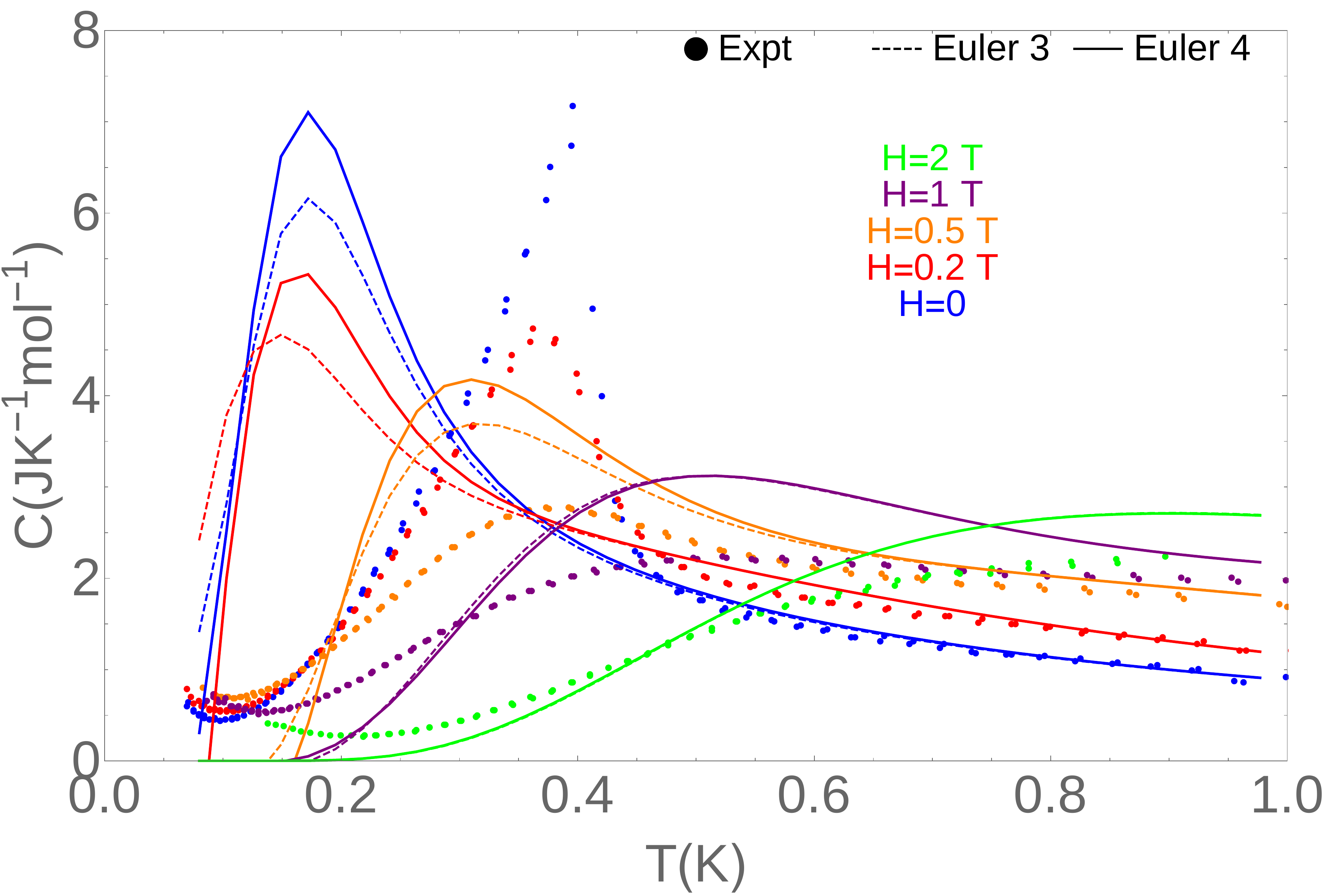}
\caption{Heat capacity in the presence of magnetic field applied along the [111] crystal axis, compared between experimental data and numerical linked cluster calculations. The NLCE calculations are performed up to fourth order, using the refined exchange parameters. The NLC expansion oscillates strongly and therefore Euler transformation is used to improve convergence. Dashed and solid lines respectively show third and fourth order NLC results with the Euler transformation. The model calculation captures the main features of the experimental data: in particular the qualitative evolution of the maximum which goes from a sharp peak for $H = 0$ and 0.2\,T to a broad smooth Schottky peak for $H = 0.5, 1.0$ and 2.0\,T as seen in experiment. However the agreement with experiment is not quantitatively good below $T\approx0.5$\,K, which may be a consequence of growing correlations extending beyond the largest cluster used in NLCE.}
\label{fig:fit_cp_all}
\end{figure}

We have used the model parameterisation to calculate the heat capacity in a magnetic field oriented along the \mbox{[111]} direction using fourth order NLCE. The terms in the series expansion oscillate strongly as the temperature is decreased, but convergence is improved by using Euler transformations \cite{Applegate2012}. Fig.~\ref{fig:fit_cp_all} shows the comparison between calculations and the experimental data at external fields of $H= 0, 0.2, 0.5, 1, 2$\,T.

The calculations at $H= 0, 0.2$\,T show a sharp increase of the heat capacity as the temperature is lowered, consistent with the approach to a phase transition. As the phase transition at $T\approx0.4$\,K is approached, the agreement between NLCE calculation and experiment becomes worse, which is an expected consequence of the correlation length exceeding the size of the largest cluster used in NLCE. For higher fields $H = 0.5, 1, 2$\,T, there is smooth crossover from high to low temperature in both calculation and experiment. Thus the NLCE calculation reproduces the qualitative behaviour seen in experiment for all fields.

\section{\label{sec:MC}Monte Carlo phase diagram}

In this section we present results for the magnetic phase diagram in an applied magnetic field determined using classical Monte Carlo simulations of the Hamiltonian determined above. We consider three directions of magnetic field: along the \mbox{[100]}, \mbox{[110]} and \mbox{[111]} crystal axes. All-in-all-out order is detected using the order parameter
\begin{equation}
m_\text{AIAO} = \frac{1}{N}\sum \tilde{\tau}_i^{\tilde z},
\label{eq:ord_par}
\end{equation}
where $N=65536$ which is the number of the spins in the simulation.

\begin{figure*}[!hbt]
\centering
\includegraphics[width=\textwidth]{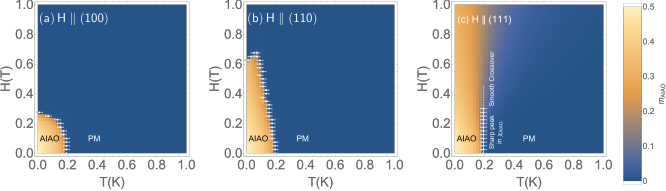}
\caption{Magnetic phase diagrams for different directions of applied magnetic field as determined from classical Monte Carlo simulations in ``field cooled'' conditions. Color scale indicates the value of the AIAO order parameter, and the white points indicate the position of a sharp peak in the corresponding order parameter susceptibility. Simulations were performed using the refined Hamiltonian parameters and a cubic cluster of $N=65536$ sites.}
\label{fig:mc_phase_dia}
\end{figure*}

\begin{figure*}[!hbt]
\centering
\includegraphics[width=\textwidth]{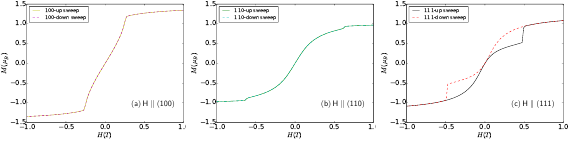}
\caption{Magnetisation curves from classical Monte Carlo simulations at $T=0.05$\,K for fields along the [100], [110] and [111] directions. On the ``down'' sweep the system is intialized from a state which is fully polarized along the direction of the field. The system is then equilibriated with $T=0.05$\,K and $H=1.0$\,T. After this magnetisation is then calculated in simulation. The final configuration in the simulation is then used to initialise the simulation at the next (lower) value of field. The ``up'' sweep works in an analogous fashion, starting from the oppositely polarised state and the opposite direction of field. For fields along the [100] and [110] directions the magnetisation curves obtained for the up and down sweeps are identical. By contrast, for fields along the [111] direction there is a large hysteresis loop.}
\label{fig:mc_mh}
\end{figure*}

The phase diagrams as a function of field strength $H$ and temperature $T$ are shown in Fig.~\ref{fig:mc_phase_dia} where the value of the order parameter in the simulation is indicated by the color scale. The phase transition temperature for each value of $H$ is estimated by the position of a sharp peak in the susceptibility of $m_\text{AIAO}$. For fields along the \mbox{[111]} direction, simulations show a sharp peak in the order parameter susceptibility for fields up to $H\sim0.3$\,T, with the peak temperature being essentially field independent. For $H>0.3$\,T, there is a smooth crossover into the low temperature phase. This is consistent with the experimental heat capacity data which shows that the sharp behavior at low field turns into a smooth crossover somewhere between 0.2\,T and 0.5\,T (Fig.~\ref{fig:fit_cp_all}).

For fields along the \mbox{[100]} direction, simulations show a sharp phase transition, with the transition temperature going to $T=0$ at $\sim0.25$\,T. For fields larger than this there is no phase transition. The measured magnetization and susceptibility in Refs.~\cite{Lhotel2015,Opherden2017} indicate a qualitatively similar behavior but with the phase transition vanishing at the lower $H_\text{c}\sim0.1$\,T.

For fields along the \mbox{[110]} direction, simulations again show a sharp phase transition, with the transition temperature going to zero at 0.6\,T. There is actually a small region of re-entrance, where there are two phase transitions for fields slightly above 0.6\,T. It would appear, based on the neutron scattering data in a \mbox{[110]} field in Ref.~\cite{Xu2018} that experimentally the AIAO order is removed by a field 0.25\,T, therefore the classical simulations overestimate the stability of the low temperature ordered state against a \mbox{[110]} field.

Finally, we consider the issue of hysteresis for the \mbox{[111]} field. We have studied the magnetisation curves at $T=0.05$\,K, by initializing the system from a polarised state and then equilibriating at $H=1$\,T. The final configuration in the simulation is then used to initialise the simulation at the next (lower) value of field. In this way the simulation is able to mimic the history dependence of a hysteresis loop and we can compare the magnetisation curves obtained when sweeping the field upwards and downwards. For fields along the \mbox{[110]} and \mbox{[100]} directions [Figs.~\ref{fig:mc_mh}(a) and (b)] there is no difference between the up and down sweeps, and thus no hysteresis. For fields along the \mbox{[111]} direction [Figs.~\ref{fig:mc_mh}(c)] there is a large hysteresis loop which we will explain in the next section. This result compares reasonably well with experiments which also show pronounced hysteresis for fields along the \mbox{[111]} direction, but only very weak hysteresis for fields along the \mbox{[100]} direction \cite{Opherden2017}.

\section{\label{sec:domain}Hysteresis and domain dependence in \mbox{[111]} fields}

\begin{figure}
\includegraphics[width=0.4\linewidth]{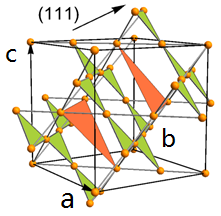}
\caption{Pyrochlore lattice decomposed into alternating triangular (orange) and kagome (green) planes stacking along the [111] directions. The golden spheres and the grey lines are the atoms and the nearest-neighbor bonds, respectively.}
\label{fig:pyro_kagome}
\end{figure}

The magnetization and AC susceptibility in magnetic fields along the \mbox{[111]} direction show abrupt changes as a function of the field, which were associated with a field-induced transition and changes in the domain structure \cite{Lhotel2015,Opherden2017}. Here we present the microscopic mechanism of the field-induced transition which reveals the existence of metastable states and domain dependence behaviors of the system in \mbox{[111]} fields.

\begin{figure}
\includegraphics[width=\linewidth]{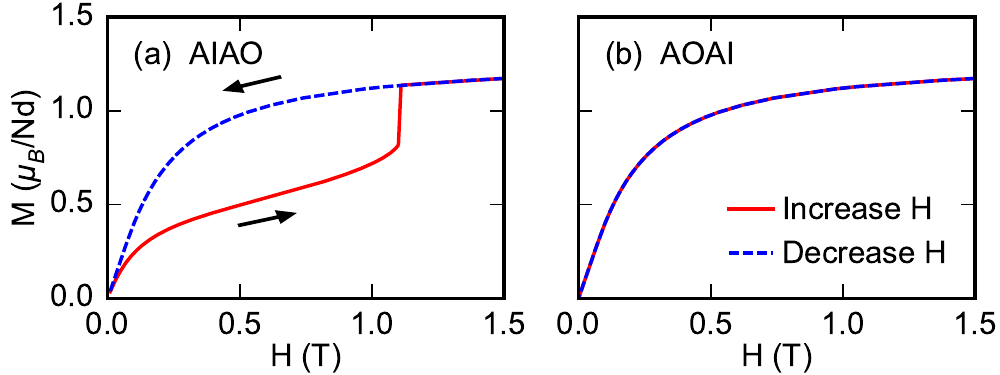}
\includegraphics[width=\linewidth]{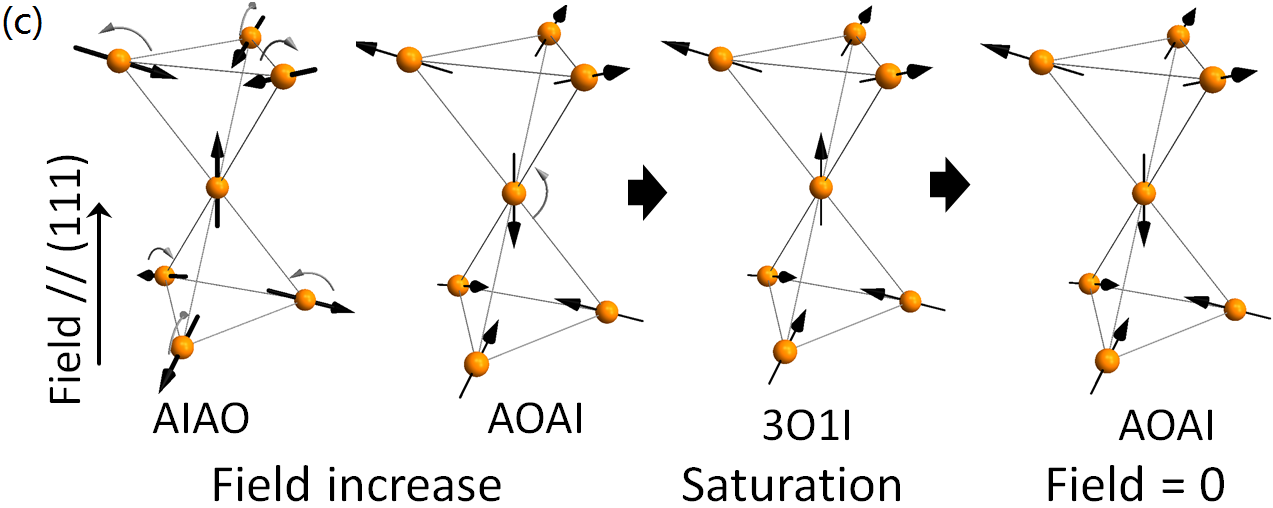}
\caption{Domain-dependent response to the [111] field of the AIAO order. (a) and (b): Calculated magnetization in increasing and decreasing fields at zero temperature for the AIAO (a) and AOAI (b) domains. (c): Schematic illustration of the spin rotation when the field is increased and then decreased.}
\label{fig:domain_mh}
\end{figure}

First, the pyrochlore lattice can be viewed as stacking of kagome and triangle planes along the \mbox{[111]} direction as shown in Fig.~\ref{fig:pyro_kagome}. Second, there are two degenerate all-in-all-out orders, namely all-in-all-out (AIAO) and all-out-all-in (AOAI), leading to the possibility for two types of magnetic domain in a sample cooled in zero field. The two types of domains are not equivalent with respect to a field along the \mbox{[111]} direction. As shown in Fig.~\ref{fig:domain_mh}, a \mbox{[111]} field flips the moments on the kagome lattice for the AIAO domain but flips the moments on the triangular lattice for the AOAI domain.

In mean field theory, assuming site-independent ansatz within the two types of layers in \mbox{[111]} fields at zero temperature, the energy on the pyrochlore lattice can be simplified to be based on a single tetrahedron given by
\begin{multline}
\varepsilon = \big[3 g_{zz} H \tau\cos(\vartheta+\phi_1)\cos\alpha + g_{zz} H \tau\cos(\vartheta+\phi_2)\cos\beta\big] \\
+6\big[\tjx\tau^2 \sin\phi_1\sin\phi_2+\tjz\tau^2\cos\phi_1\cos\phi_2\\
+\tjx\tau^2\sin^2\phi_1 + \tjz\tau^2\cos^2\phi_1\big]
\label{eq:mf_e}
\end{multline}
where the spins are taken as classical vectors $\tau[\cos(\phi_i),0,\sin(\phi_i)]$ with $\tau=1/2$ and $\phi_1$ and $\phi_2$ are the spin canting angles with respect to the ${\tilde z}$ axes in the ${\tilde x}$-${\tilde z}$ planes for the kagome and triangular lattices and $\cos\alpha=1/3$ and $\cos\beta=1$ with $\alpha$ and $\beta$ being the angles between the local [111] axes and the [111] field for the kagome and triangular lattices. Here we have only two angular variables for the magnetic lattice with four sublattices because there only two types of sites (in the kagome and triangle planes) in the \mbox{[111]} field. The non-magnetic $\tilde{\tau}^{\tilde{y}}$ component is excluded for the current situation because $\tau_{\tilde y}$ is not coupled to the field directly ($g_{yy}=0$) and $\tilde{J}_{\tilde y}$ is small compared with the other terms.

Fig.~\ref{fig:domain_mh}(a) and (b) shows the calculated magnetisation in increasing and decreasing \mbox{[111]} fields obtained by optimizing the energy (Eq.~\ref{eq:mf_e}) locally for the two types of domains (the optimized state in the previous field is used as the start for the current field). The calculation is consistent qualitatively with the susceptibility data in Ref.~\cite{Opherden2017} and the Monte Carlo simulations in Sec.~\ref{sec:MC}. We found that the hysteresis only happens to the AIAO domain. Fig.~\ref{fig:domain_mh}(c) shows the moment configurations in the progress of changing the field for the AIAO and AOAI domains. It reveals that the flip of the moments on the kagome layers for the AIAO domain causes the sudden jump in the magnetization and in decreasing field the moments on the triangular layers reverse their direction smoothly recovering to an AOAI state (not the initial AIAO state) leading to the irreversibility. On the contrary, the AOAI domain always has the moments on the triangular layers flipped and reversed smoothly for increasing and decreasing fields. This result is consistent with the recent single crystal neutron diffraction experiments in magnetic fields \cite{Lhotel2018} and supports the proposed domain inversion explanation for the field dependence of the magneto-resistivity in the pyrochlores containing 5$d$ elements \cite{Tardif2015,Fujita2016,Tian2016,Fujita2018,Ma2015}.

\begin{figure}
\includegraphics[width=\linewidth]{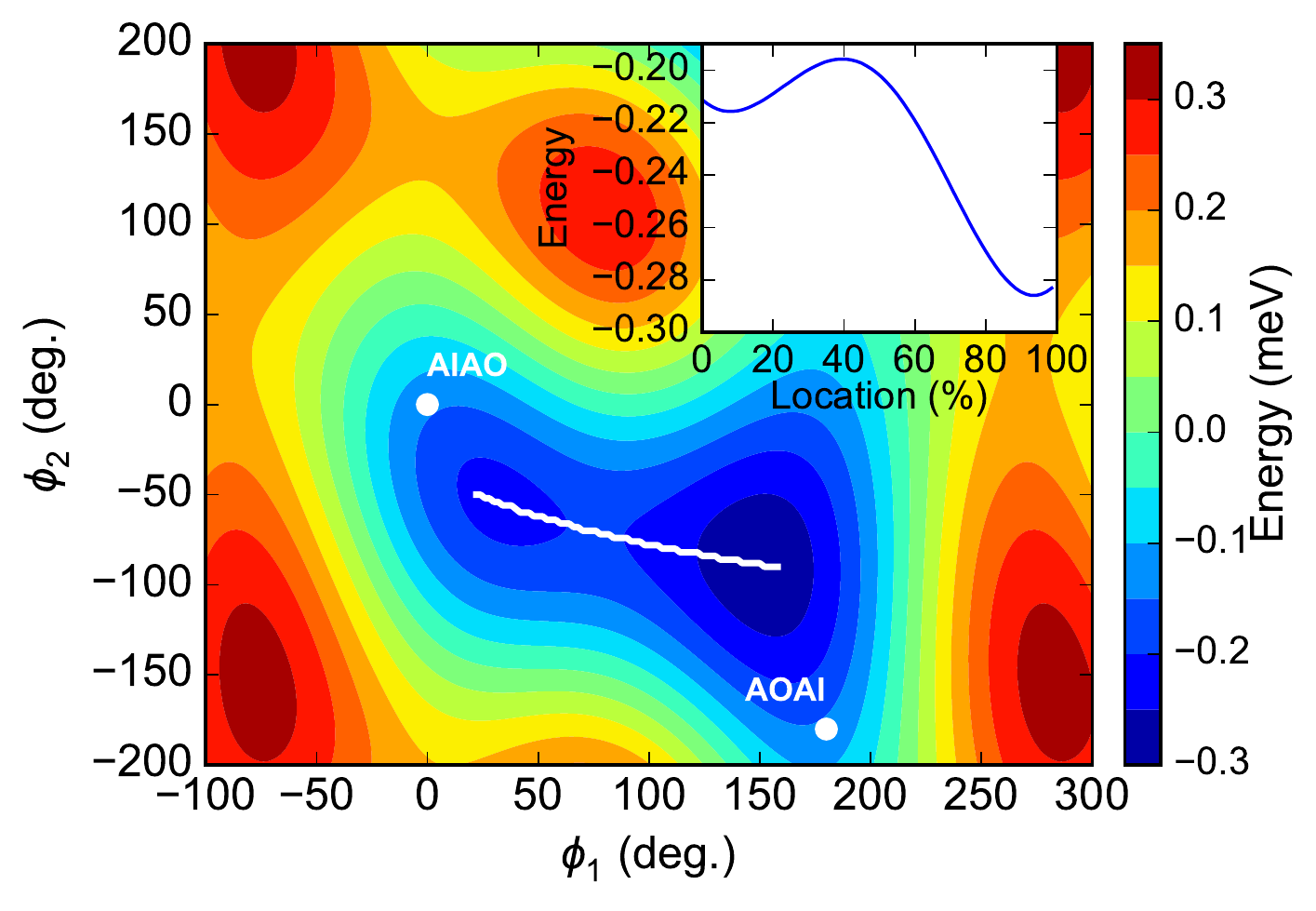}
\caption{Energy landscape calculated based on Eq.~\ref{eq:mf_e} for the AIAO order in the 0.8~T [111] field as a function of the spin rotation angles on the kagome and triangle lattices. The inset shows the energy along the cut (the white line in the main figure) through the two minima where the energy barrier is clearly shown.}
\label{fig:domain_e}
\end{figure}

To investigate the irreversibility further, we calculated the energy landscape for field along the \mbox{[111]} direction. Fig.~\ref{fig:domain_e} shows the energy landscape as a function of the spin orientations on the kagome and triangular layers in the [111] field of 0.8\,T (lower than the mean-field $H_\text{c}\sim1.2$\,T) with the exchange Hamiltonian determined above. As we can see, the AOAI spin configuration is smoothly connected to the global energy minimum corresponding to a field polarized 3-in-1-out state and there is an energy barrier in-between for the AIAO domain. Therefore the AIAO state is a metastable state in the [111] field. In an increasing field, when the energy barrier vanishes at the critical field, the AIAO state transforms to the ground state leading to a sudden change of the spin configuration. When decreasing the field, the polarized state returns to AOAI state which is smoothly connected to the polarized state in energy.

The recovery to the AOAI single-domain state from the field-polarized 3-in-1-out state also can be understood by considering the exchange fields on the kagome and triangle layers. In the field-polarized state, the spins on the triangle layers are subjected to a stronger molecular field against the applied magnetic field because they have six broken bonds with their neighbors from the kagome layers while the spins on the kagome layers only have two.

\begin{figure}
\includegraphics[width=\linewidth]{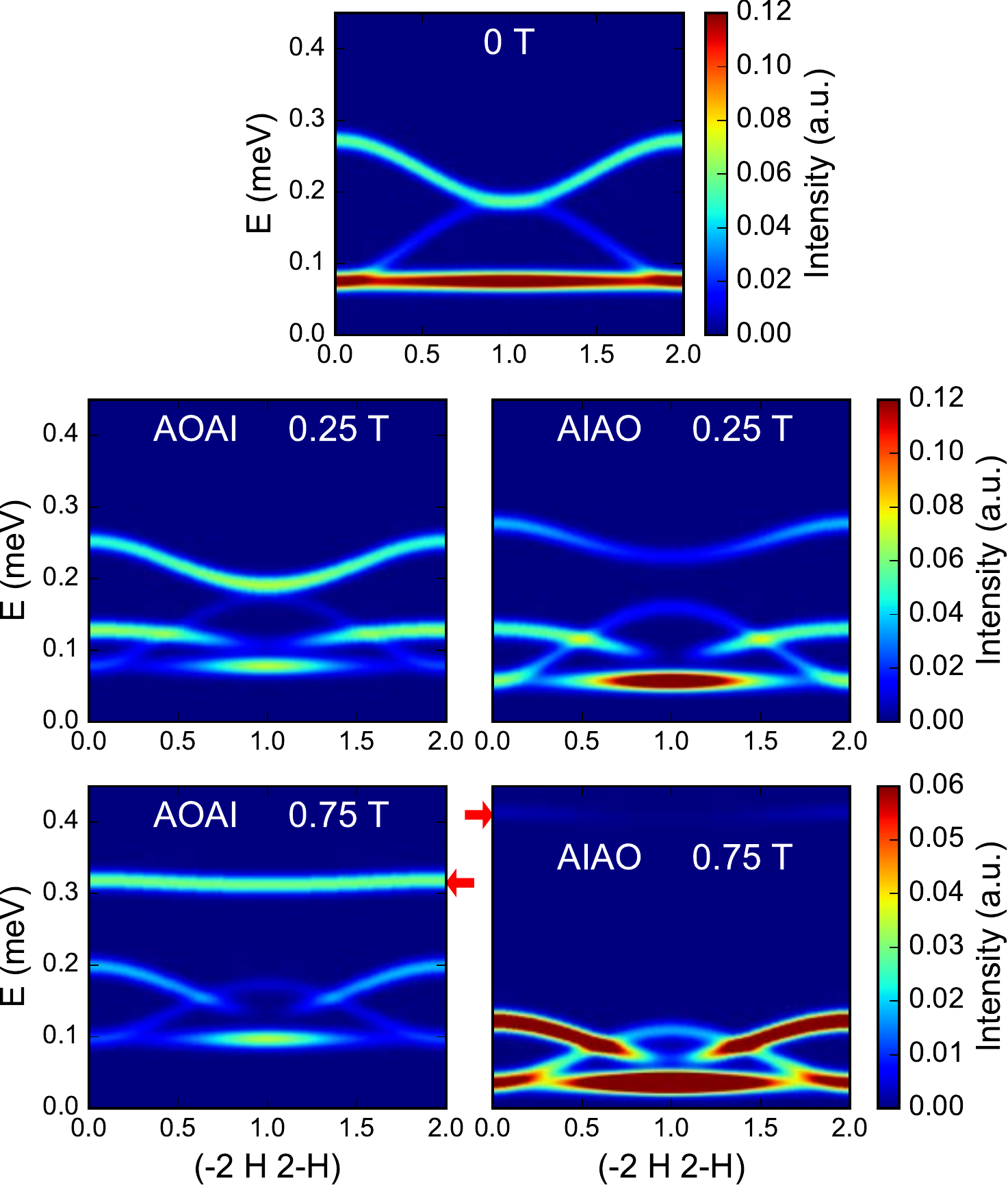}
\caption{Calculated neutron scattering spectra for the AOAI (left) and AIAO (right) domains in [111] fields. The red arrows indicates the position of the flattening modes.}
\label{fig:domain_spinw}
\end{figure}

The domain-dependent response to the \mbox{[111]} field has significance for the spin dynamics, i.e. dynamical kagome spin ice observed in \ndzro\ in [111] fields \cite{Lhotel2018}. First, the two domains have different spin configurations in fields below the critical field and are expected to show different spin dynamics. Our spin wave calculation shows that the gap to the dynamical kagome ice modes exhibits opposite behaviors for the two types of domains: decreasing for the AIAO domain and increasing for the AOAI domain. This may contribute to the broadening of the line width in the INS data \cite{Lhotel2018} if the sample is cooled in zero field and contains two types of domains. Second, above $H_\text{c}$ only the AOAI domain exists in the sample based on which the spin wave simulation should be done specifically though the AIAO domain is still present in mean-field theory. As shown in the Fig.~\ref{fig:domain_spinw}, the spin wave of the AOAI domain shows a better fit to the data in Ref.~\cite{Lhotel2018} (the experimental $H_\text{c}\sim0.1$\,T). However, there are still discrepancies especially in scattering intensity between the data and theory which is attributed to the limitation of mean field theory and the possible quantum fluctuations in the system. Third, it was shown that the two-dimensional dynamical kagome ice is quite robust appearing both below and above the critical field \cite{Lhotel2018}. This is because the spin configuration on the kagome lattice is always 3-in-3-out in \mbox{[111]} fields.

\section{\label{summary}Conclusions}

We refined the spin Hamiltonian by the combined analyses of the high-energy-resolution inelastic neutron scattering data, susceptibility, magnetisation and specific heat of \ndzro\ using linear spin wave theory and numerical linked cluster expansion. Using classical Monte Carlo simulation, we have calculated the magnetic phase diagram in the presence of applied magnetic fields in the \mbox{[100]}, \mbox{[110]} and \mbox{[111]} directions which qualitatively agrees with the experimental phase diagram and reproduce the large hysteresis seen for fields along \mbox{[111]}. However, simulations overestimate the stability of the low temperature all-in-all-out order against external magnetic field, compared to experimental results. This may be related to quantum effects as shown in Yb$_2$Ti$_2$O$_7$ \cite{Hitesh2017}.

The microscopic mechanism for the hysteresis is revealed in our mean-field calculation to be the existence of a metastable state and the AIAO-AOAI domain inversion in [111] field. Including the domain-dependent response to the \mbox{[111]} field, the observed dynamical kagome ice can be well described though there are still disagreements \cite{Lhotel2018}. In addition, the microscopic mechanism for the domain inversion could also be applied to the Nd-containing 5$d$ pyrochlores which shows hysteresis in the magneto-resistivity as a function of the field \cite{Tardif2015,Fujita2016,Tian2016,Fujita2018,Ma2015} and may also be related to the anomaly in the AC susceptibility of Nd$_2$Hf$_2$O$_7$ \cite{Opherden2018}.

\begin{acknowledgements}
We thank Y.-P. Huang, M. Hermele, S. T. Bramwell and A. T. Boothroyd for helpful discussions on the related theory.  We acknowledge Helmholtz Gemeinschaft for funding via the Helmholtz Virtual Institute (Project No. VH-VI-521). This research used resources at the Spallation Neutron Source, a DOE Office of Science User Facility operated by the Oak Ridge National Laboratory. Experiments at the ISIS Neutron and Muon Source were supported by a beamtime allocation RB1810504 from the Science and Technology Facilities Council (DOI: 10.5286/ISIS.E.92924095).
\end{acknowledgements}


\appendix

\section{\label{appendix:xrd}X-ray diffraction and crystallography}
\begin{figure}[!htb]
\centering
\includegraphics[width=0.8\linewidth]{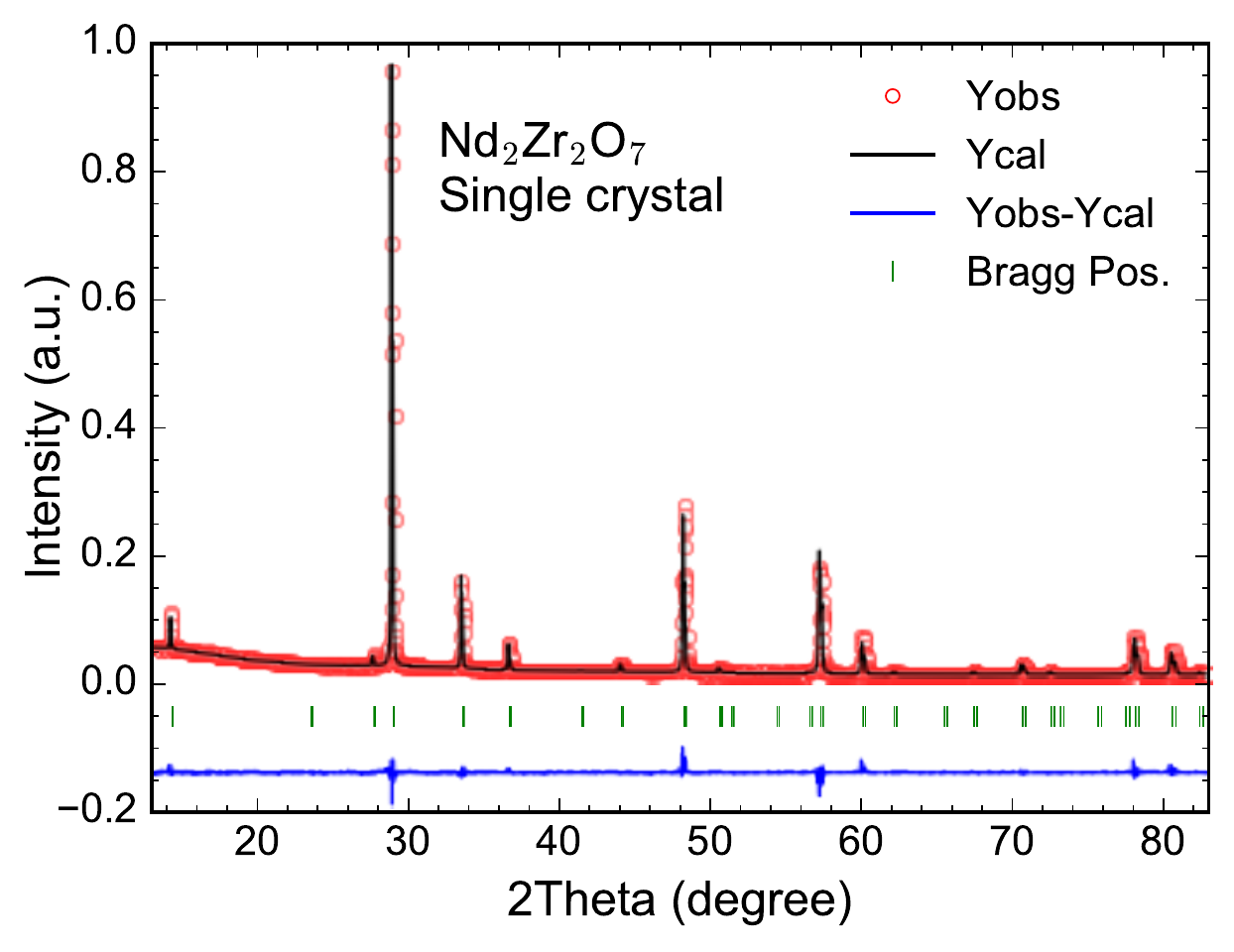}
\caption{\label{fig:xrd} X-ray powder diffraction pattern of crushed single crystal \ndzro\ at room temperature where the  observed (Yobs), calculated (Ycal), Yobs-Ycal and the Bragg peak positions (Bragg Pos) are shown.}
\end{figure}

The structure of the single crystal was characterized using powder X-ray diffraction (XRD) (Bruker-D8, Cu-$K_\alpha$) and Rietveld refinements using the software Fullprof Suite \cite{fullprof}. The XRD pattern and refinement are shown in Fig.~\ref{fig:xrd}. The crystallographic parameters are $a=10.651(1)$\,\AA\, and $x_\text{O1}=0.3347(1)$ which are consistent with the previous reports for the powder and single crystal samples \cite{Xu2015,Lhotel2015,Hatnean2015}

\section{\label{appendix:chi_mag}Susceptibility and magnetization}
\begin{figure}[!htb]
\centering
\includegraphics[width=0.8\linewidth]{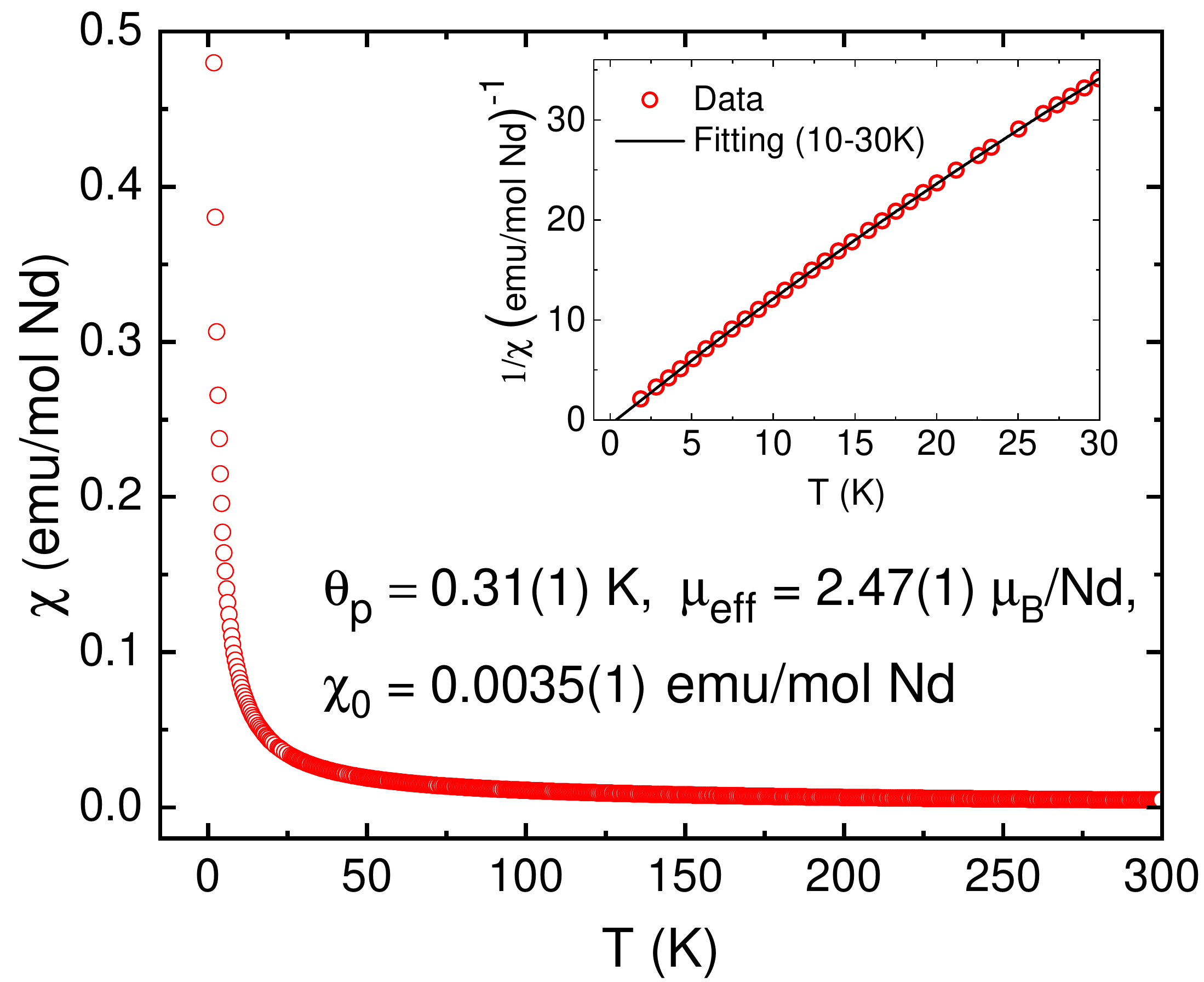}
\caption{Susceptibility of the single crystal \ndzro\ measured with field 0.1\,T applied along the [100] direction and the Curie-Weiss law fitting (inset). Demagnetizing effect is corrected.}
\label{fig:nd_dc_chi}
\end{figure}

Figure.~\ref{fig:nd_dc_chi} shows the susceptibility of the single crystal sample of \ndzro\ above 2\,K. The susceptibility increases with decreasing temperature and shows no anomaly above 2~K. The $\chi(T)$ data were fitted by the Curie-Weiss law with the Van Vleck term $\chi(T) = \chi_0 + C/(T-T_\text{cw})$ for the temperature range 10\,$\leq T\leq 30$\,K. This yields $T_\text{CW}=0.31(1)$\,K, $\mu_\text{eff}=2.47(1)\,$\mub/Nd and $\chi_0=3.5(1)\times10^{-3}\,$emu/mol\,Nd, which are consistent with those reported for powder and single crystal samples \cite{Xu2015,Lhotel2015,Hatnean2015}. The positive $T_\text{CW}$ indicates an effective ferromagnetic interaction between the \ndt\ moments though the sample shows an ``all-in-all-out'' antiferromagnetic order. This is a consequence of the dipolar-octupolar nature of the effective spin of the \ndt\ ion in pyrochlores \cite{Benton2016}.

\begin{figure}[!htb]
\centering
\includegraphics[width=0.8\linewidth]{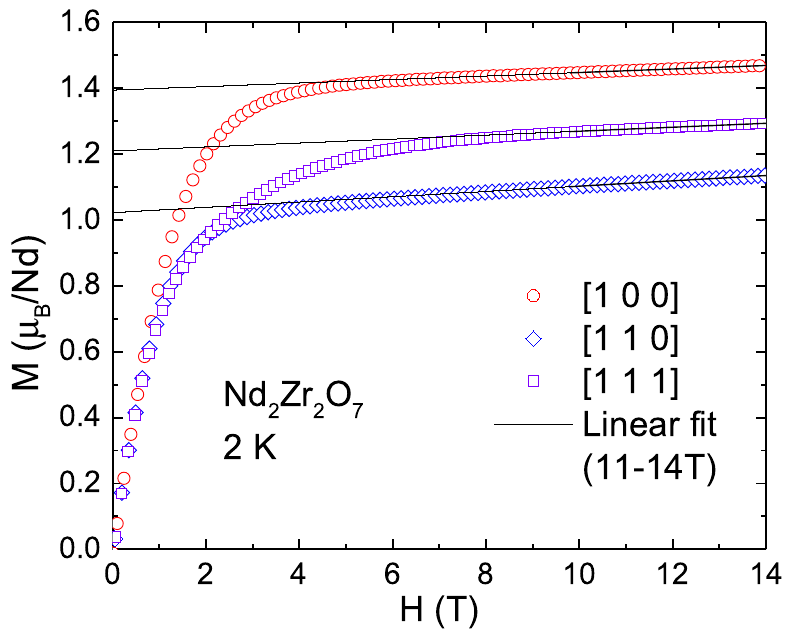}
\caption{Magnetization of the single crystal at 2\,K with fields along different directions. The black line shows the linear fitting for the Van Vleck contribution in high fields.}
\label{fig:nd_dc_m}
\end{figure}

Figure.~\ref{fig:nd_dc_m} shows the magnetization $M(H)$ data for the single crystal with field along the three main cubic directions at 2\,K. The magnetization is highly reduced (comparable to the free-ion value) and anisotropic, consistent with the previous reports \cite{Lhotel2015,Hatnean2015}. The magnetizations with fields applied along the [100] and the \mbox{[110]} directions are the highest and the lowest, respectively, reminiscent of the spin ice materials, consistent with the Ising anisotropy of the \ndt\ moment found in the crystal field analyses \cite{Xu2015, Lhotel2015}.

After subtracting the Van Vleck susceptibility, the data shows saturated magnetizations $M_\text{s}$ 1.39, 1.02 and 1.21\,\mub/Nd for the three directions. For a pyrochlore with local \mbox{[111]} Ising anisotropy, $M_\text{s}$ for the three directions should be $\mu(1/\sqrt 3)$ , $\mu(\sqrt{2/3}\times2)/4$ and $\mu(1+1/3\times 3)/4$, respectively where $\mu$ is the magnetic moment of \ndt\ \cite{Fukazawa2002}. With the refined $\mu=g_{zz}\mu_\text{B}/2=2.5(1)\,$\mub, we have 1.44(5) 1.02(4) and 1.25(5)\,\mub\ which is consistent with the data within 5\%\ deviation.

\section{\label{appendix:cp}Specific heat}
\begin{figure}[!htb]
\centering
\includegraphics[width=0.8\linewidth]{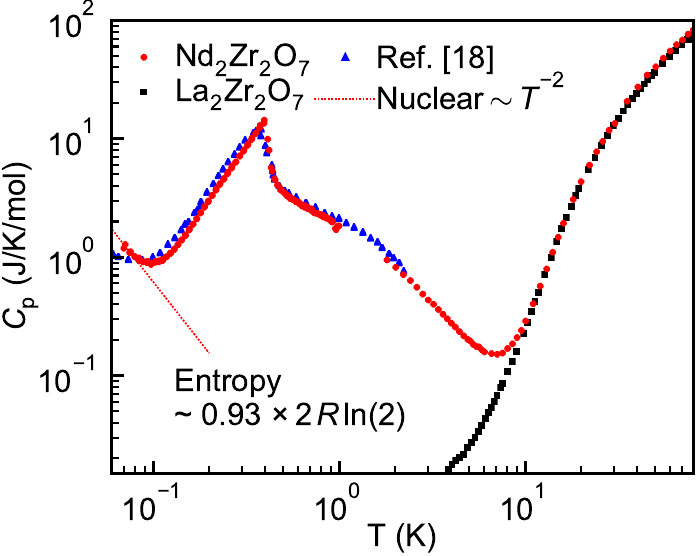}
\includegraphics[width=0.9\linewidth]{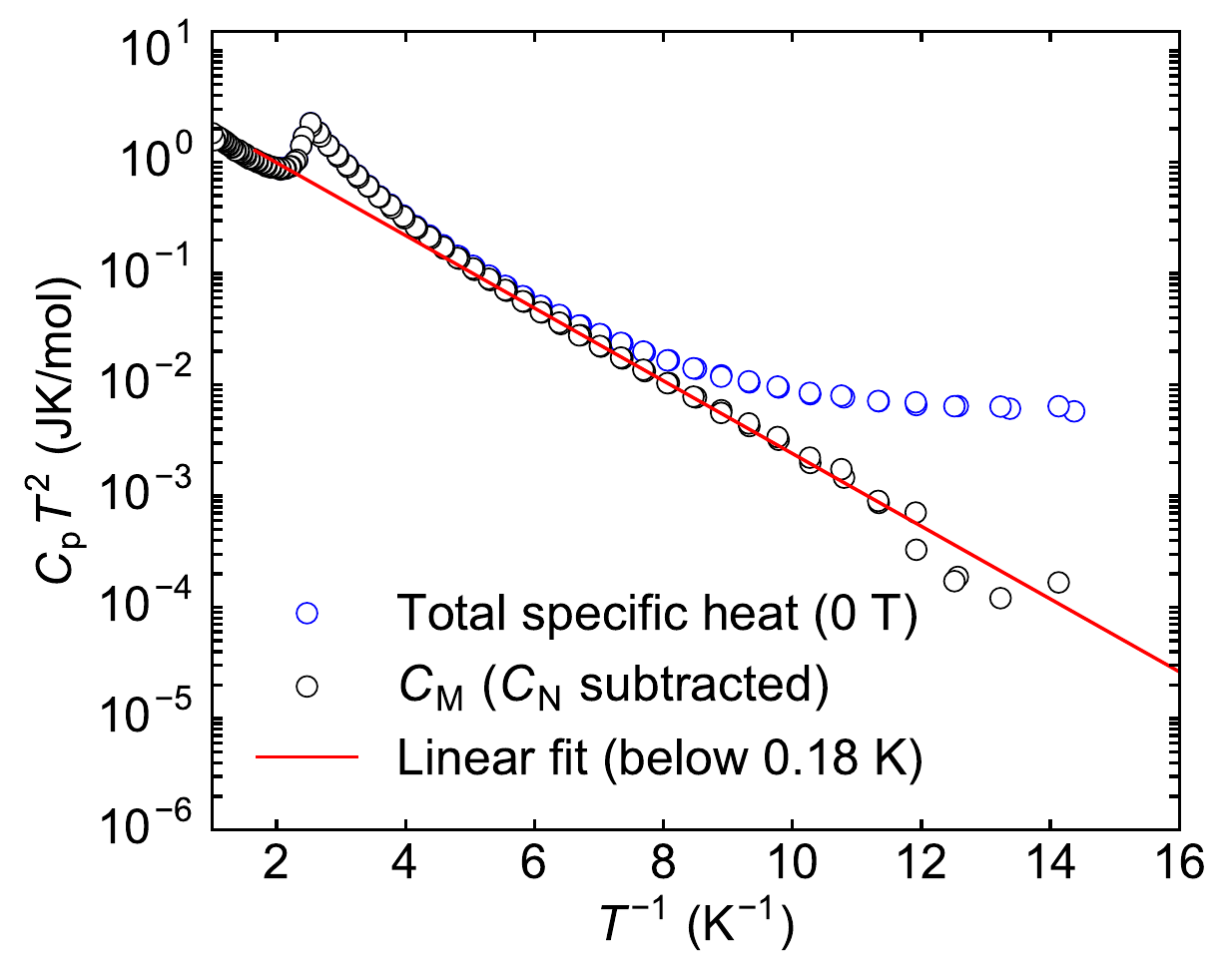}
\caption{(upper panel) Specific heat of single crystal \ndzro\ and powder \lazro\ samples. The magnetic entropy is calculated based on the data below 10\,K after subtracting the fitted nuclear part ($\sim T^{-2}$) and the measured phonon part (the \cp\ of \lazro). (lower panel) Fit to the data below 0.18\,K with $A\,T^{-2} + B\,T^{-2} \exp(-\Delta/T)$ which shows the raw data, the magnetic 4$f$ electron part $C_\text{M}$ obtained by subtracting the fitted nuclear part $C_\text{N}$ from the raw data and the fit.}
\label{fig:nd_cp_lowT}
\end{figure}

Figure~\ref{fig:nd_cp_lowT} shows the specific heat \cp\ data of single crystal \ndzro\ and powder \lazro\ samples and the data for powder \ndzro\ in Refs.~\cite{Blote1969,Lutique2003}. Above 10\,K, the specific heat of the two compounds are nearly the same because of their similar structures and thus the \lazro\ data was used as a non-magnetic background (at higher temperatures they are different due to the CEF effect) \cite{Xu2015}. At low temperatures, \ndzro\ shows a $\lambda$-shape peak at \tn$\,\approx0.4\,$K similar to the powder sample \cite{Blote1969,Lutique2003}. The upturn below 0.1\,K is caused by the nuclear hyperfine interactions whose contribution is $\sim T^{-2}$ in the high temperature region \cite{Blote1969}. The magnetic 4$f$ electrons contribute to the \cp\ peak, which originates from the magnetic correlations and excitations.

After subtracting the phonon and nuclear contributions, the calculated magnetic entropy is $\sim0.93\,R\ln(2)$/mol\,Nd below 10\,K. The entropy released due to the phase transition is close to $R\ln(2)$/mol\,Nd, indicating the establishment of long-range order in a spin-1/2 system, consistent with the neutron diffraction and crystal field analysis in Refs.~\cite{Lhotel2015,Xu2015}.

For a normal antiferromagnet with a linear magnon dispersion at low energy, the \cp$(T)$ should show a $T^{3}$ temperature dependence at sufficient low temperature below \tn. It was reported that the $T^3$ law applies to the low temperature \cp\ of \ndsno\ which has the same magnetic order as for \ndzro\ \cite{Bertin2015,Xu2015}. However, the inelastic neutron scattering data of \ndzro\ clearly shows gapped magnon excitations. Therefore, the $T^3$ law is invalid for \ndzro. The model $\sim T^{-2}\exp(-\Delta/T)$ ($\Delta$ is the magnon gap size) is normally used to describe the temperature dependence of the specific heat of gapped magnon excitations \cite{Quilliam2007}. It would be a straight line in the $\log(C_\text{p}T^2)$-$1/T$ plot. The data below 0.18\,K is fitted to $A\,T^{-2} + B\,T^{-2} \exp(-\Delta/T)$ (Fig.~\ref{fig:nd_cp_lowT}) which yields $\Delta=0.75(3)\,$K. The obtained gap is quite close to the measured one ($\approx0.075\,\text{meV}\,=\,0.87\,\text{K}$) as shown in Sec.~\ref{sec:spinH} (see Ref.~\cite{Xu2017} for details).

\section{\label{appendix:sw_fit}Spin wave fitting}
\begin{figure*}[!hbt]
\centering
\includegraphics[width=\textwidth]{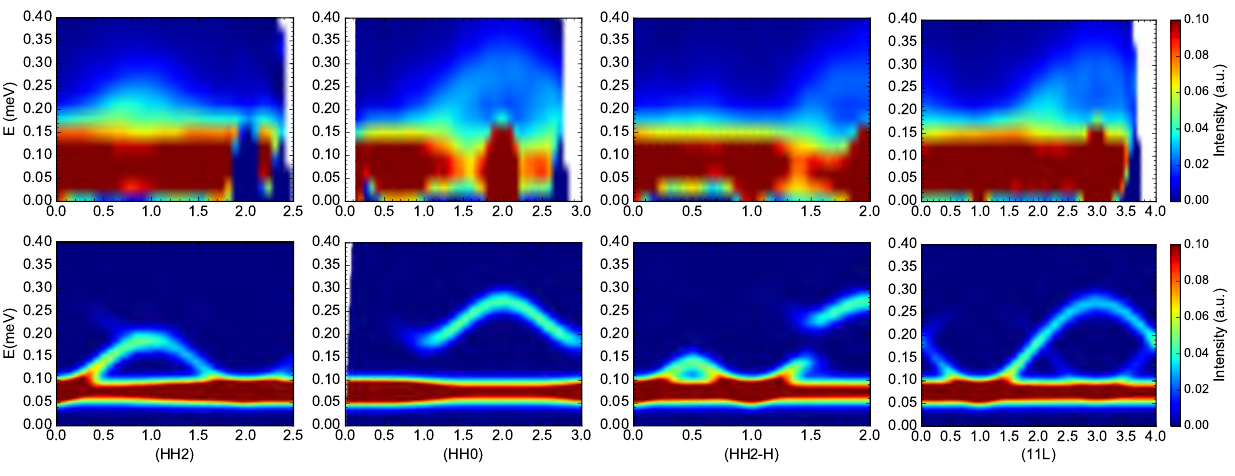}
\caption{(first row) Inelastic neutron scattering spectra (in arb. units) taken at 240\,mK along different high symmetry directions presented by color-coded intensity $E-Q$ maps. The background has been subtracted using the high temperature (20\,K) dataset. (second row) Corresponding spin wave calculation based on the pseudospin 1/2 model for the ordered phase, using the exchange parameters in Eq.~\ref{eq:j_par}. The experimental resolution is 0.1\,meV while 0.04\,meV resolution is used in the calculation in order to show the dispersion clearly. The calculated pattern is normalized to the data by an overall factor for better comparison.}
\label{fig:ins_spinw_1}
\end{figure*}

\begin{figure*}[!hbt]
\centering
\includegraphics[width=\textwidth]{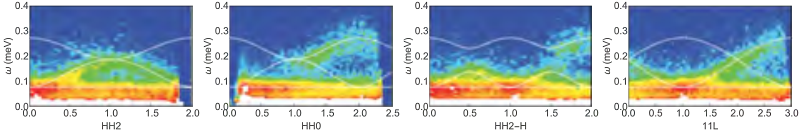}
\caption{Spin wave dispersion above the all-in-all-out ground state calculated from the refined exchange parameters (white dashed lines), plotted over the inelastic neutron scattering results from Ref.~\cite{Petit2016}.}
\label{fig:fit_Petit_data}
\end{figure*}

Figure~\ref{fig:ins_spinw_1} shows the INS data measured on CNCS comparing with the calculated scatering pattern using linear spin wave theory. Fig.~\ref{fig:fit_Petit_data} shows the magnon dispersion plotted over the INS data of Ref.~\cite{Petit2016}.

\section{Numerical Linked Cluster Expansion (NLCE)}
\label{appendix:NLCE}

\begin{figure*}[!hbt]
\centering
\includegraphics[width=0.6\textwidth]{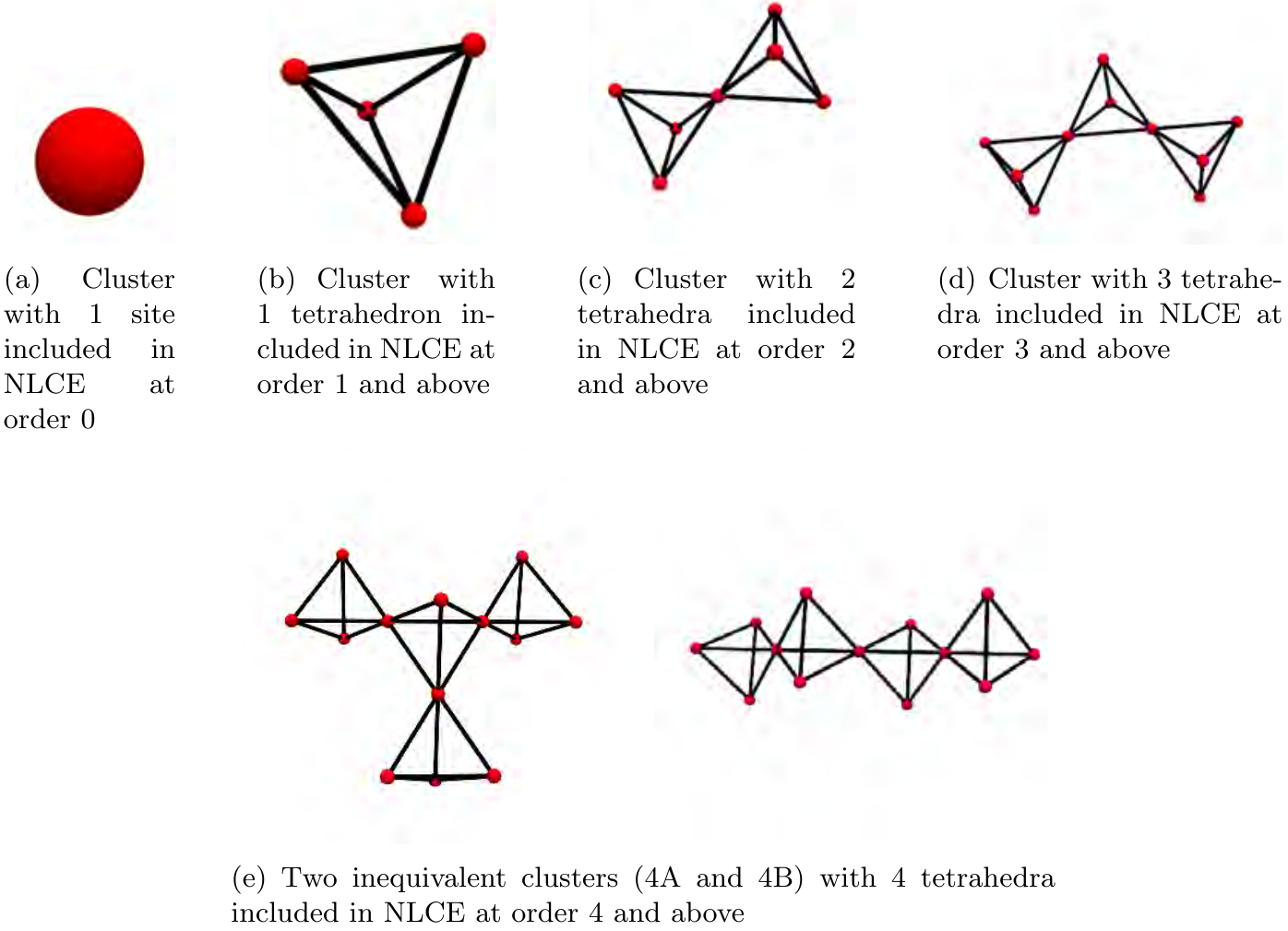}
\caption{Clusters used in NLCE calculation. The $n^\text{th}$ order of the NLCE calculation includes clusters of up to $n$ tetrheadra. Here we have taken the calculation up to fourth order.}
\label{fig:clusters_nlce}
\end{figure*}

Figure~\ref{fig:clusters_nlce} shows the clusters used in the NLCE calculation. We have used Numerical Linked Cluster Expansion (NLCE) to
calculate the magnetic susceptibility and heat capacity for
comparison with experiment. An introduction to the NLCE method is given in [\onlinecite{Tang2013}].

NLCE calculates extensive quantities per site  $\frac{\langle \mathcal{O} \rangle}{N}$ as sums over contributions from clusters $c$ which can be embedded in the lattice
\begin{eqnarray}
\frac{1}{N}\langle \mathcal{O} \rangle = \sum_{c } M(c) W(c).
\label{eq:NLCsum}
\end{eqnarray}
$M(c)$ is the number of times $c$
can be embedded in the lattice, divided by the number of sites $N$. $W(c)$ is the weight  of the cluster:
\begin{eqnarray}
W(c)=\langle \mathcal{O} \rangle_c-\sum_{s \subset c} W(s)
\label{eq:NLCweight}
\end{eqnarray}
where $\langle \mathcal{O} \rangle_c$ is the expectation value of
$\mathcal{O}$ calculated from exact diagonalization of cluster $c$ with open boundary conditions. The second term in Eq. (\ref{eq:NLCweight}) is a sum over the weights of all subclusters
of $c$.

We use a series of clusters beginning with a single site (zeroth order NLC, NLC0) and
then all subsequent clusters are constructed from full tetrahedra. The $n^\text{th}$ order of NLC (NLCn), calculates the sum in Eq.~(\ref{eq:NLCsum}) up to clusters of $n$ tetrahedra. Our calculations of the magnetic susceptibility (Fig.~\ref{fig:fit_ht_chi}) were made up to third order (NLC3) and calculations of the heat capacity (Fig.~\ref{fig:fit_ht_chi}) were made up to fourth order (NLC4).

In the absence of a magnetic field, there is only one kind of cluster at each order up to NLC3, and two distinct clusters at NLC4. These are illustrated in Fig.~\ref{fig:clusters_nlce}. In the presence of an applied magnetic field along the $111$ direction there are two distinct types of cluster which must be taken into account in each of NLC2 and NLC3 and six distinct types of cluster in NLC4.

To improve convergence, we have used Euler transformation on the heat capacity calculations
\cite{Applegate2012}. The Euler transformed result at third order is:
\begin{eqnarray}
 \langle \mathcal{O} \rangle_{\sf Euler 3}=
 \frac{1}{2} \langle \mathcal{O} \rangle_{\sf NLC 2} + \frac{1}{2} \langle \mathcal{O} \rangle_{\sf NLC 3}
\end{eqnarray}
where $\langle \mathcal{O} \rangle_{\sf NLC n}$ is the estimate of $\langle \mathcal{O} \rangle$
up to $n^{\text{th}}$ order in NLC. The Euler transformed result at fourth order is
\begin{eqnarray}
 \langle \mathcal{O} \rangle_{\sf Euler 4}=
 \frac{1}{4} \langle \mathcal{O} \rangle_{\sf NLC 2} + \frac{1}{2} \langle \mathcal{O} \rangle_{\sf NLC 3}
+ \frac{1}{4} \langle \mathcal{O} \rangle_{\sf NLC 4}
\end{eqnarray}

\section{Monte Carlo Simulations}
\label{appendix:MC}

Classical Monte Carlo simulations have been used to calculate
the magnetic phase diagrams in applied field (Fig.~\ref{fig:mc_phase_dia}) and hysteresis curves (Fig.~\ref{fig:mc_mh}), for the [111], [110] and [100] field directions. In these simulations the spins are treated as classical vectors with fixed length $|{\bf S}|=1/2$. The simulations use the standard Metropolis algorithm, with single spin updates using the Marsaglia method \cite{marsaglia72}. Simulations are performed on cubic clusters of $N=16 L^3$ sites, with $L$ being the number of 16-site cubic unit cells along each side of the cluster. We define one Monte Carlo step (MCS) as one sweep of the whole
lattice of $N$ sites, attempting an update at each site.

Details of the simulations for the phase diagrams are given in Appendix \ref{appendix:MC_pd} and for the hysteresis loops in Appendix \ref{appendix:MC_hysteresis}.

\subsection{Determination of phase diagrams}
\label{appendix:MC_pd}

The phase diagrams in Fig.~\ref{fig:mc_phase_dia} were determined
using simulations on an $L=16$ ($N=65536$) cluster. The simulations were performed separately for each value of applied field and followed the following protocol:
\begin{enumerate}
\item{The system was initalized to a random spin configuration.}
\item{The system was then equilibriated at $T=1$\,K using 5000\,MCS
without measurements being taken.}
\item{Observables are then averaged over the course of 50000\,MCS,
with each observable measured every 50\,MCS.}
\item{The temperature is then reduced by $\delta T=0.0202$\,K, and the system
is equilibriated at the new temperature using 500 MCS.}
\item{The cycle is repeated down to $T=0.01$\,K.}
\end{enumerate}

\subsection{Hysteresis}
\label{appendix:MC_hysteresis}

The magnetic curves in Fig.~\ref{fig:mc_mh} were determined using simulations on an $L=4$ ($N=1024$) cluster. The simulations for the down sweep followed the following protocol:
\begin{enumerate}
\item{The system was initalized to a spin configuration fully polarized along the direction of the field.}
\item{The system was then equilibriated
with an external field of $H=1.0$\,T and  temperature $T=0.05$\,K
using 500\,MCS.}
\item{Observables are then averaged over the course of 100000\,MCS,
measuring every 100\,MCS.}
\item{The magnetic field is then decreased by $\delta H=0.0202$\,T, and the system
is equilibriated at the new value of external field using 500\,MCS.}\\
\item{The cycle is repeated down to $H=-1.0$\,T.}
\end{enumerate}

The simulations for the up sweep are exactly the same, but initialising from
a negatively polarised state and increasing the field from $H=-1.0$\,T.

\bibliographystyle{abbrv}

\end{document}